\documentclass[12pt,fleqn]{article}
\usepackage{amsfonts,amsmath,amssymb,graphicx,setspace,natbib}
\usepackage[usenames, dvipsnames]{xcolor}
\usepackage[colorlinks=true,linkcolor=blue,citecolor=Mahogany,urlcolor=blue]{hyperref}
\usepackage[top=1in, bottom=1in, left=1in, right=1in]{geometry}
\usepackage[normalem]{ulem}
\usepackage{pdflscape}
\usepackage{booktabs}
\usepackage{graphicx}
\usepackage{array}
\usepackage{subcaption} 
\newtheorem{theorem}{Theorem}
\newtheorem{proposition}[theorem]{Proposition}

\newtheorem{assumption}{Assumption}

\usepackage{multirow} 

\begin{document}

\title{{\Large \textbf{Ignorance Is Bliss: \\ The Screening Effect of (Noisy) Information}}\thanks{Previously titled ``Looking the Other Way: The Screening Role of (Weak) Internal Monitoring". We thank Judson Caskey (the editor) and two anonymous reviewers, Phil Dybvig, Paolo Fulghieri, Kebin Ma, Beatrice Michaeli, Sebastian Pfeil, Richard Saouma, Chao Tang, Curtis Taylor, Alexei Tchistyi, Andrew Winton, Huaxiang Yin, and participants at the MIT Asia Conference in Accounting, Summer Institute of Finance, Global Management Accounting Research Symposium, Econometric Society NASM, Stony Brook ICGT, the Conference on Markets and Economies with Information Frictions, the Financial Markets and Corporate Governance Conference, RSM Corporate Finance Day, CFAR, TJAR, and EEA-ESMS for helpful comments and suggestions. Robin Luo provided excellent research assistance.}}
\author{Felix Zhiyu Feng\thanks{Michael G. Foster School of Business, University of Washington. Email: ffeng@uw.edu} \quad \quad Wenyu Wang\thanks{Kelley School of Business, Indiana University. Email: wenywang@indiana.edu}\quad\quad Yufeng Wu\thanks{Fisher College of Business, Ohio State University. Email: wu.6251@osu.edu}\quad\quad Gaoqing Zhang\thanks{Carlson School of Management, University of Minnesota, and Tepper School of Business, Carnegie Mellon University. Email: zhangg@umn.edu}}

\date{{\large \textit{The Accounting Review, forthcoming}}}
\maketitle

\begin{abstract}
\begin{spacing}{1}
\noindent This paper studies the value of a firm's internal information when the firm faces an adverse selection problem arising from unobservable managerial abilities. While more precise information allows the firm to make \textit{ex post} more efficient investment decisions, noisier information has an \textit{ex ante} screening effect that allows the firm to attract on-average better managers. The trade-off between more effective screening of managers and more informed investment implies a non-monotonic relationship between firm value and information quality. A marginal improvement in information quality does not necessarily lead to an overall improvement in firm value.


\end{spacing}
\end{abstract}

\newpage

\section{Introduction}
Internal information system and managerial talent both play critical roles in firm operation in practice. Firms rely on high-quality information to assess the prospect of their investment opportunities, and on high-quality managers to safeguard the success of any investments undertaken. While the literature has long recognized the importance of information quality and managerial screening, these two are usually studied separately and in distinct settings. 

In this paper, we propose and analyze a model in which the two decisions interact with each other. Consistent with the conventional wisdom, a firm makes more efficient investment decisions when internal information is more precise. However, we demonstrate the existence of a novel channel in which information quality also affects the screening of managers whose ability is unobservable. Moreover, we find that surprisingly, the firm achieves a more effective screening result from \textit{noisier} information. The trade-off between better screening and more informed investment decisions results in a non-monotonic relationship between the quality of a firm’s information system and its overall value: an improvement in the former does not always lead to an increase in the latter. 

In the model, a firm needs to hire a manager from a pool of potential candidates to supervise a risky investment project. Each candidate's innate ability, which we refer to as his \textit{type}, determines the probability that the project will succeed if he is hired. To hire a manager, the firm offers a compensation contract to all candidates, but faces an adverse selection problem because it does not observe each candidate's type. As a result, the contract must offer the managers a certain amount of utility, known as the managers' \textit{rent}, in exchange for the truthful reporting of their types. Seeing the contract, each manager forms rational expectations of the rent he can extract based on his own type. Those who are willing to accept the firm's contract form a managerial pool, from which a randomly selected manager will be hired. We normalize the value of all managers' outside options to zero, and assume they prefer unemployed when they are indifferent. Therefore, the managerial pool consists of only the candidates who can expect to receive positive rent from the contract. 

After the project manager is hired but before the investment is made, the firm receives a signal from its internal information system about whether the project will succeed or not. The more precise the signal is, the more likely a good (bad) signal indicates eventual success (failure). The firm can utilize this signal to make the decision of whether to continue or scrap the project. Continuation incurs costs for both the firm and the manager, but generates some output only if the project succeeds. If the project is scrapped, both the firm and the manager receive zero payoff. Thus, the managerial pool is equivalent to the set of managers who expect to receive investment under some conditions. 

Crucially, while the investment outcome is assumed to be observable and contractible, the firm's signal is not. This reflects the fact that internal information regarding a firm's operation is usually soft, subjective, and difficult to fully characterize in a formal contract. Consequently, the firm cannot \textit{ex ante} commit to any investment policy contingent on the realization of the signal. Instead, by calculating the \textit{ex post} conditional probability of success based on the manager's (reported) type and the actual signal received, the firm makes one of three investment decisions: never invest, always invest, and invest only when the good signal is received. This divides the space of the manager candidates into three investment zones. Low-type managers with insufficient conditional probability of success regardless of the signal are in the \textit{no investment} zone. High-type managers with sufficient conditional probability of success even under the bad signal are in the \textit{unconditional investment} zone. The remaining intermediate-type managers are in the \textit{conditional investment} zone, within which investment is made if and only if the firm receives a good signal. 

The range of the conditional investment zone depends on the quality of the signal. When the signal is more precise, the conditional probability of success given the good signal is higher for any type of manager. Consequently, the conditional investment zone expands, and the minimal type of manager that can expect to receive investment becomes lower. Meanwhile, because the manager's rent comes from his compensation, which he can receive only if investment is made, the minimal type of manager in the conditional investment zone is also the minimal type in the managerial pool. This implies our main result that a more precise signal reduces the average quality of the manager the firm can hire, which we refer to as the \textit{screening effect} of information quality. When information is near perfect, i.e., when a good (bad) signal leads to success (failure) almost surely, the investment decision is made based on the signal only. All managers are in the conditional investment zone, and no manager will be excluded from the managerial pool. In contrast, when the signal is noisier, the firm relies less on the signal to make the investment decision. The conditional investment zone shrinks while the average quality of the managerial pool increases. In the limiting case when the signal is completely uninformative, the conditional investment zone disappears, and the investment decision is solely based on the type of the manager. Only the managers with sufficiently high types (i.e., sufficiently high unconditional likelihood of success) will receive investment. All other managers are excluded from the managerial pool, allowing the firm to achieve the best screening result.

Next, we analyze the overall effect of information quality on the \textit{ex ante} firm value, i.e., the expected firm profit before hiring the manager. While a noisier signal achieves more effective screening, it does not necessarily imply that the firm is always better off. In the model, the firm has two objectives: hiring better managers, and investing only when it is likely going to succeed. Noisier information helps the firm achieve the first objective, but better information helps the firm achieve the second objective. We show this trade-off of the dual roles of information leads to a non-monotonic relationship between information quality and firm value. For firms with a strong existing information system, the effect of more informed investment dominates, and firm value increases in information quality. However, for firms with a weak existing information system, the screening effect of noisier information dominates, and firm value actually decreases in information quality. 

Our model produces several implications for the value of internal information system in practice. In particular, our results suggest that \textit{marginal} improvements of information quality are not always beneficial, and the overall impact of small or gradual adjustments of a firm's internal information system on firm value should be evaluated carefully and comprehensively. Indeed, some practitioners and researchers have urged firms to use caution when obtaining information about investment activities of their managers (\citealp*{Bernstein2014,DeCremer2016,McKinsey2017,Hearn2020}). Interestingly, these articles note that too much information may have a negative impact that makes talented managers unwilling to work for the firms. We formalize and extend this intuition via a model that involves adverse selection and limited commitment, and quantitatively characterize the conditions under which the averse effect of better information may dominate its positive effect of directing investment more efficiently. Our results also provide a possible explanation for the observed heterogeneity among firms' internal governance practices, such as the manners and intensities at which firms conduct internal monitoring or auditing, or the various degrees of latitude the managers enjoy in making operational decisions. These diverse policies could be the result of firms balancing the direct effect of more informed decisions and the indirect effect of screening more capable managers. In particular, a lack of frequent monitoring and auditing or a large degree of discretion for the managers is not necessarily the result of failure in internal governance.

\section{Related Literature}

Our paper is most closely related to the literature that explores the impact of internal information on firms' real and financial decisions and their implications on firm value (e.g., \citealp*{Mehran1992, Hermalin2005,  BurnsKediaLipson2010, BaldeniusMelumadMeng2014}, etc.), while highlighting a potential cost associated with more precise information. There are two groups of studies that, like ours, point to the downside of better information, although our study focuses on a different mechanism. The first group of studies consider moral hazard as the main friction and the principal's lack of commitment power as a critical constraint. For example, \citet*{arya1997interaction} consider a double-moral hazard problem in which both the principal and the agent can exert costly effort to influence the outcome of a risky production technology. The principal can be better off not generating the additional information, allowing her to credibly commit to low effort and reducing the incentive payments needed for the agent. Such adverse effect can likewise be found when better information results in excessive forgiveness (\citealp{cremer1995arm}),  relaxed budget constraint (\citealp{segal1998monopoly}), or the lack of financial discipline (\citealp{dewatripont1995credit}). Meanwhile, \citet{Zhu22} demonstrates an opposite example in which more information can cause the principal to be too harsh, using the information for excessive punishment. Similar to these studies, the value of less information in our model also stems from its ability to alleviate the principal's limited commitment constraint. Different from them, however, we study an adverse selection problem arising from unobservable managerial type (as opposed to the moral hazard problem studied in \citealp*{arya1997interaction}). The firm cannot commit not to use its signal to make investment decision, which attracts rent-seeking low-type managers and lowers the average quality of the managerial pool. In that regard, the focus of our paper is the role of information design in the screening of managers whereas \citet*{arya1997interaction} focus on the issue of alleviating moral hazard problems. Moreover, while obtaining high-quality information is costly in some of the previous studies (e.g., \citealp{cremer1995arm} and \citealp{Zhu22}), the quality of information is intrinsically costless in our model, but carries two endogenous effects (screening versus informed investment) that manifest in opposite directions. 

The second group of studies on the potential cost of better information also considers adverse selection as part of the frictions but assumes the informed agent has multiple actions. Better information can mitigate the agency friction in one action but exacerbate the friction in the other action. For example, \citet{meng2020board} consider a setting in which the manager must procure information about the quality of an investment project and implement the project. While a board with higher expertise can better align its assessment of the project with that of the manager, the board's expertise also increases the cost of providing incentives for the manager to truthfully report his assessment and undertake the desired implementation effort. In similar spirits, \citet*{BGP97}, \citet*{kanodia2005imprecision}, \citet{GS06}, and \citet{BS21} assume that the agent possesses private information while simultaneously taking private actions, and more knowledge about either the private information or the private action raises the cost of incentives for the other.\footnote{Other studies in this category include \citet{morris2002social}, where the agents have multiple objectives, and \citet{goldstein2019good}, where decision makers face multiple sources of uncertainty.} Our paper shows that even though the manager has a single role of reporting private information, internal information itself has multiple countervailing effects. In particular, noisier information can facilitate the \textit{ex ante} screening of managers when the managers' ability is not directly observable.

Our paper is also related to the earnings management literature to the extent that earnings management yields less precise accounting reports. Our analysis implies that managed earnings can be more desirable than unmanaged earnings, considering the screening benefit of less precise information systems. In that regard, our paper is connected to studies highlighting the positive side of earnings management (e.g., \citealp*{arya1998earnings,arya2008performance,gao2018reporting}, etc.). In our model, the benefit of less precise information also stems from its value as a commitment device. However, unlike the existing studies, the value of coarser information in our paper manifests through the \textit{ex ante} screening of managers that effectively limits the amount of rent low-type managers can extract from the firm.

Finally, our paper also contributes to the literature on job assignments, where adverse selection and its resolution is a primary concern. Prior work on this topic emphasizes the role of task-specific skills (e.g., \citealp{gibbons2004task,lazear2009firm,lise2020multidimensional,tervio2008difference}) or firm characteristics, such as size or culture, that affect the firm-labor matching (e.g., \citealp{gabaix2008has,rivera2012hiring}). Our paper complements these studies by highlighting the screening role of internal information about firms' investment projects in this market.

\section{Model}
This section introduces the model set to capture the basic process of a firm hiring \textit{a project manager}, and the role of the firm's information system in screening the pool of candidate managers. Section \ref{subsection:timing} lays out the timing of the events and the basic model structure. Section \ref{subsection:assumptions} discusses and motivates the key assumptions. Section \ref{subsection:objectives} describes the optimization objective of each party and defines the equilibrium.

\subsection{Timing of Events} \label{subsection:timing}

A firm needs to hire a manager from a pool of potential candidates to supervise a risky investment project. Both the firm and the managers are risk-neutral, but the latter are protected by limited liability. Figure \ref{fig:time} summarizes the timing of events, which are described in detail below:

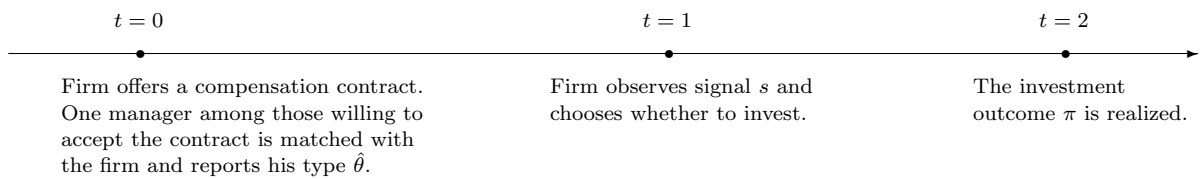
\begin{figure}[h]
\bigskip
\bigskip	
{\normalsize 
\begin{picture}(50,70)(-75,-65)
\put(-70,0){\vector(1,0){450}}

\put(-30,10){\scriptsize{$t=0$}}
\put(-20,0){\circle*{3}}

\put(170,10){\scriptsize{$t=1$}}
\put(180,0){\circle*{3}}

\put(320,10){\scriptsize{$t=2$}}
\put(330,0){\circle*{3}}

\put(-50,-15){\scriptsize{Firm offers a compensation contract.}}
\put(-50,-25){\scriptsize{One manager among those willing to}}
\put(-50,-35){\scriptsize{accept the contract is matched with}}
\put(-50,-45){\scriptsize{the firm and reports his type $\hat{\theta}$.}}

\put(135,-15){\scriptsize{Firm observes signal $s$ and}}
\put(135,-25){\scriptsize{chooses whether to invest.}}

\put(295,-15){\scriptsize{The investment}}
\put(295,-25){\scriptsize{outcome $\pi$ is realized.}}

\end{picture}}
\caption{\medskip Timeline of the model.}
\label{fig:time}
\end{figure}

At $t=0$, the firm is endowed with an opportunity to invest in a risky project. The payoff of the project $\pi \in \left\{ 0,1\right\} $ is binary: either the project succeeds and returns a unit of output, or it fails and yields $0$. The success of the project requires the firm to hire a manager to supervise the operation of the project, and the success probability $\theta \in \lbrack 0,1]$ depends on the manager's ability (type). The pool of potential manager candidates is a continuum indexed by each manager's type $\theta $, which is privately known only to the manager himself. All the other parties share a common belief that $\theta $ follows a distribution characterized by a cumulative density function $F(\theta)$.

The firm offers a compensation contract in order to hire a manager from the candidate pool. For simplicity, we assume that the firm is cash-constrained and, therefore, cannot make payments to the manager unless the project succeeds and generates cash flows. Accordingly, the firm's compensation contract to the manager is a wage $w\geq 0$ payable if and only if the project succeeds.\footnote{Such compensation contract offered to the manager can be micro-founded in a variant of our model augmented with a moral hazard problem of the manager. We formally analyze such a model in Appendix \ref{appendix:MH}. Our analyses suggest that, to solicit effort from the manager, the firm offers an equilibrium compensation contract that pays the manager a bonus only upon project success. Furthermore, the level of the bonus is the same as the equilibrium wage $w$ derived in our baseline model. Accordingly, all of our subsequent analyses remain intact with the addition of the moral hazard problem.} We allow the firm to optimally tailor its wage structure. In particular, it can be made contingent on the manager's reported type.

We normalize the managers' outside option to zero, and assume that they prefer to remain unemployed when they are indifferent. Therefore, managers are willing to accept the firm's contract if their expected net payoff from working in the firm is strictly positive.\footnote{Alternatively, we can assume the managers prefer to be employed when indifferent, but the value of their outside option is strictly positive. We present the analysis of this extension in Appendix \ref{appendix:outside option}. The results are qualitatively the same.} For simplicity, we assume a one-time random matching between the firm and the pool of managers willing to accept the firm's contract. Upon his employment, the manager reports his type $\hat{\theta}$. We place no restrictions on the manager's reporting choice.

At $t=1$, the firm receives a binary signal $s\in\{0,1\}$ from its internal information system regarding the outcome $\pi$ of its project. The quality of the signal is captured by a pair of parameters $\{q_1, q_0\}$, where
\begin{align}
\Pr (s=0|\pi =1) & = q_1, \\
\Pr (s=1|\pi =0) & = q_0. 
\end{align}%
In other words, $\{q_1, q_0\}$ represent the probability that the signal is the opposite of the underlying state. A higher $q_1$ or $q_0$ thus represents a \textit{noisier} information system. We assume that $\{q_1, q_0\}\in(0,1/2]$, so that $s=1$ can be appropriately labeled as the good signal and $s=0$ as the bad signal based on the implied likelihood of the high and low states, respectively, where $q_1=q_0=1/2$ represents the case that the signals are fully uninformative.\footnote{Note that we have ruled out the case that the signal $s$ reveals the investment outcome perfectly, i.e., $q_0 = q_1 = 0$. The perfect-information case is uninteresting for our analysis because in this case, the firm does not screen managers at the hiring stage. To see this, note that when $s$ is fully informative, the firm invests if and only if $s$ reveals perfectly an outcome of success. Accordingly, all managers, regardless of their types, receive non-negative rents and join the managerial pool as long as the firm offers a wage upon success $w$ that at least covers the manager's private investment cost $c$.} We assume that the signal is the firm's private information and is unverifiable, but their quality $\{q_1, q_0\}$ is publicly known. 

Upon observing the signal and conditional on the manager's reported type $\hat{\theta}$, the firm chooses whether to invest in its risky project. If the firm chooses not to invest, both the firm and its manager receive zero payoffs. Alternatively, if the firm chooses to invest, it incurs an investment cost $k>0$. Investment also incurs a cost $c>0$ from the manager regardless of whether it eventually succeeds. We assume that $k+c<1$ to ensure that the net surplus from the project is positive if it succeeds.

If the investment is made, the outcome is realized at $t=2$, and payments are made according to the firm's contract. 

\subsection{Discussion of Assumptions} \label{subsection:assumptions}

We now discuss and motivate several key assumptions.

We model the firm's signal $s$ to capture the firm's (such as its board of directors') preliminary assessment of the project's prospect after minimal work has been conducted to lift the project ``off the ground." Accordingly, we impose the following assumptions on the properties of the signal $s$. First, because it is reasonable to imagine that a project cannot meaningfully get started without a dedicated team that includes a supervisor or coordinator (i.e., the project manager), the signal is only available after the manager is hired.\footnote{Alternatively, this can be understood as the board's assessment requires insights or some basic information gathered by the project manager, e.g., \citet{meng2020board}.} Meanwhile, as soon as the project is in motion, some information regarding its potential to succeed becomes available before additional investment of resources is required. Thus, the signal is available before the production decision that involves the costs $k$ and $c$, which can be regarded as much more substantial than that needed for the project's commencement. Second, the assessment naturally involves soft and subjective information, such as the board of directors' own evaluation, interpretation, and inferences, which are difficult to be formally defined or legally described. Thus, we assume that the signal $s$ is private and unverifiable so that the firm cannot \textit{ex ante} commit to an investment strategy based on its realization. However, because the assessment of a project prospect requires specialized skills, the managerial candidates can develop knowledge/expectations about the precision of such assessment based on observable firm characteristics, such as the composition, background, industry expertise, and experience of the board members. Thus, we assume that the pool of candidate managers observe the quality of the signal, $q_1$ and $q_0$, at $t=0$, and then formulate rational expectations about their payoffs if they accept the firm's contract given the firm's information quality. Such assumption regarding the common knowledge of the information quality is also made in the extant literature, e.g.,  \citet*{kamenica2011bayesian}, \citet*{michaeli2017divide}, \citet{meng2020board}, \citet*{bertomeu2021strategic}, \citet*{friedman2020optimal,friedman2022rationale}, \citet*{dordzhieva2022signaling}, etc.

We assume the manager incurs a personal cost $c$ as long as investment is made. In other words, to focus on the screening mechanism in the presence of adverse selection, we do not model hidden effort choices by the manager or any resulting moral hazard problem. In Appendix \ref{appendix:MH}, we relax this assumption and demonstrate the robustness of our results if investment is also subject to moral hazard. 

Lastly, we assume a one-time random matching between the firm and the pool of manager candidates willing to accept the firm's contract. This is mainly to simplify the structure of the managerial labor market. It rules out the competition among managers and any potential repeated signaling or bargaining games. However, the fact that each candidate can decide whether to join the managerial pool after seeing the contract is crucial. As we explain later, this allows the firm to use the contract terms to shape the distribution of managers that it can potentially be matched with, i.e., the contract serves as a screening device for the firm. The assumption can also be relaxed, for example, by allowing some sorting between the manager and the firm, which we discuss in Appendix \ref{appendix:rand match}. In sum, as long as every manager who joins the managerial pool has some probability of being matched with the firm, our mechanism remains qualitatively intact.

\subsection{Optimization Objectives and Equilibrium} \label{subsection:objectives}

Because type is the manager's private information, the firm faces an adverse selection problem. The firm's compensation contract and investment policy must, therefore, provide each manager with the incentives to report his type truthfully. Formally, let $p_{s}(\hat{\theta})\equiv \Pr (\pi =1|s,\hat{\theta})$ denote the conditional probability that the project will succeed given the signal $s$ and the manager's reported type $\hat{\theta}$, and $w(\hat{\theta})$ denote the wage to the manager at success. Because the firm cannot \textit{ex ante} commit, its investment decision, denoted by $x(s,\hat{\theta})$, must be \textit{ex post} optimal, i.e., 
\begin{equation} \label{eq:optimal x}
x(s,\hat{\theta})= 
\begin{cases}
1,\text{ if }p_{s}(\hat{\theta})[1-w(\hat{\theta})]>k, \\ 
0,\text{ otherwise.} 
\end{cases}%
\end{equation}%
Given this investment policy, each manager forms rational expectations of the likelihood of receiving investment and chooses his reporting strategy $\hat{\theta}$ to maximize his expected utility from the contract, denoted by 
\begin{equation} 
R_{\hat{\theta}}(\theta )=\mathrm{E}[x(s,\hat{\theta})(w(\hat{\theta})-c)].
\end{equation}%
In our later analysis, we focus on incentive-compatible (IC) contracts under which each manager finds it optimal to report his type truthfully, i.e.,
\begin{equation} \label{eq:IC}
\theta =\hat{\theta}^{\ast }\equiv \arg \max_{\hat{\theta}}R_{\hat{\theta}}(\theta ),\text{ for all }\theta.
\end{equation}%
We then drop the notation $\hat{\theta}$ in the subsequent analyses as long as the IC condition is met or imposed. For example, the manager's expected utility from the contract in the equilibrium is simply denoted by $R(\theta )\equiv R_{\hat{\theta}^{\ast }=\theta }(\theta )$. This is commonly known as the \textit{managerial rent}, which refers to the utility that a manager needs to receive in exchange for revealing his private information truthfully. 

The manager will accept the contract if and only if he receives a positive rent from the firm's contract. That is, the firm's managerial pool $\Theta$ is defined as 
\begin{align}  \label{eq:PC}
\Theta = \{\theta: R(\theta)>0\}.
\end{align}

The firm's objective is to maximize its \textit{ex ante} expected payoff before matching with a manager and has two instruments (controls) to do so: wage $w$, and investment policy $x$. To save notations, define the vector $\mathbf{q} \equiv [q_1\; q_0]^\top$ and let
\begin{align}  \label{eq:V(q) definition}
V(\mathbf{q}) = \max_{x,w} \int_\Theta \mathrm{E} \left[x(s,\theta) (\pi - w(\theta)-
k)\right]dF(\theta),
\end{align}
denote the expected firm value under a given level of information quality.\footnote{We use $F(\theta)$ as the cumulative density function for the overall, unconditional distribution of $\theta$ and use $\int_\Theta v(\theta)dF(\theta)$ to represent the conditional expectation of any function $v(\theta)$ given $\theta\in\Theta$. As it will become clear later, $\Theta = \{\theta:\theta>\theta_1\}$ for some $\theta_1$ in the equilibrium. In that case, 
\begin{equation*}
\int_\Theta v(\theta)dF(\theta) = \frac{1}{1-F(\theta_1)}\int_{\theta_1}^1 v(\theta)dF(\theta).
\end{equation*}} 
Consequently, we define a firm's contract as \textit{optimal} if the contract solves (\ref{eq:V(q) definition}) subject to constraints (\ref{eq:optimal x}), (\ref{eq:IC}), and (\ref{eq:PC}). That is, the investment decisions $x$ must be \textit{ex post} optimal; the contract must be incentive compatible; and the managerial pool consists of only those with strictly positive rents. We refer to the policies and payoffs under the optimal contract as the \textit{equilibrium} policies and payoffs. 

Our focus on the truthtelling equilibrium merits some additional discussion. In short, this focus reflects the trade-off between generality and tractability. As noted earlier, constructing an equilibrium in our model consists of characterizing the following components: the firm's wage offer, the manager's reporting strategy, the firm's belief about the manager's type given the report, and the firm's investment strategy given the report, belief, wage, and the signal about the investment outcome. All components are endogenously and jointly determined in the equilibrium, formulating a fixed-point problem with multiple controls that is generally difficult to solve. Therefore, to gain better analytical tractability, one must pre-specify some structure on the types of equilibria to focus on. In the extant literature, truthtelling is one such natural choice, as it arguably yields the highest degree of economic insights.\footnote{See, for instance, \citet{athey2013efficient}, \citet{deb2015dynamic}, and \citet{doval2022mechanism} for applying the truthtelling requirements in mechanism design problems with partial commitment.} It also alleviates the challenges associated with solving our fixed-point problem with multiple controls by allowing us to solve for the equilibrium components sequentially, as we demonstrate in the next section. Nevertheless, readers who are interested in the robustness of our main insights can refer to the Online Appendix of this paper, where we show that the screening effect of noisier information holds more broadly under other plausible equilibria that do not necessarily involve truthtelling.

\section{Model Solution} \label{section:solution}

This section presents the solution to the model. We derive the equilibrium managerial rent under a given information system in Section \ref{subsection:rent} and present how our main result pertaining to the screening effect varies with the firm's information quality in Section \ref{subsection:screening}. Section \ref{subsection:role of commitment} offers an alternative explanation for the intuition of the screening effect for interested readers. 

\subsection{Managerial Rent} \label{subsection:rent}

We begin by deriving the firm's optimal screening contract (i.e., the wage and investment policy) and the implied equilibrium managerial rent for a given level of information quality. For ease of exposition, we will derive the equilibrium heuristically first, before collecting all the results to present formally. In particular, we will first conjecture that, in equilibrium, there is an incentive-compatible contract under which the wage for success is a constant, i.e., $w(\theta) = w$ for all $\theta$. We verify later that this is indeed a property of the optimal contract.

Given the firm's information quality \{$q_1, q_2$\} and the type of the manager hired $\theta$, the probability of all events can be summarized in the following matrix:

\begin{center}
\begin{tabular}{rcc}
\hline\hline
& Success $\pi= 1$ & Failure $\pi= 0$ \\ \hline
Good signal $s= 1$ & $\theta (1-q_1)$ & $(1-\theta)q_0$ \\ 
Bad signal $s= 0$ & $\theta q_1$ & $(1-\theta)(1-q_0)$ \\ \hline\hline
\end{tabular}
\end{center}
We can define $p_{s}(\theta)$, the conditional probability of success given
the signal $s$ and type $\theta $, as: 
\begin{align}
p_{1}(\theta )\equiv \Pr (\pi =1|s=1)& =\frac{\theta (1-q_1)}{\theta
(1-q_1)+(1-\theta )q_0},  \label{eq:p1} \\
p_{0}(\theta )\equiv \Pr (\pi =1|s=0)& =\frac{\theta q_1}{\theta q_1+(1-\theta
)(1-q_0)}.  \label{eq:p0}
\end{align}%
Conditional on each signal, the firm makes the investment if and only if $p_{s}(\theta)(1-w)>k$. This implies two cutoff types: $\theta_{0}^{f}$ and $\theta_{1}^{f}$, which represent the points at which the firm is indifferent between investment or not conditional on a bad signal ($s=0$) and on a good signal ($s=1$), respectively. The superscript $f$ denotes these cutoffs based on the firm's investment decision. These two cutoffs solve the two corresponding indifference conditions: 
\begin{align}
p_{1}(\theta_{1}^{f})(1-w)& =k,  \label{eq:continuation condition1} \\
p_{0}(\theta_{0}^{f})(1-w)& =k.  \label{eq:continuation condition0}
\end{align}%

It is straightforward to show that $\theta_{1}^{f}<\theta_{0}^{f}$ when $q_1, q_2 <1/2$. Intuitively, for any given $\theta $, the good signal is more likely to be associated with success than the bad signal is. Therefore, the minimal type of manager that the firm is willing to invest in is lower upon observing a good signal than a bad signal. Moreover, 
\begin{align}
\lim_{\theta \rightarrow 0}p_{0}(\theta )& =\lim_{\theta \rightarrow 0}p_{1}(\theta )=0, \\
\lim_{\theta \rightarrow 1}p_{0}(\theta )& =\lim_{\theta \rightarrow 1}p_{1}(\theta )=1,
\end{align}%
which means that as the manager's type approaches zero (one), the conditional likelihood of success approaches zero (one) regardless of the signal. Since the return from investment is strictly negative (positive) when the success likelihood is zero (one), the space of managerial type is divided into three \textquotedblleft zones\textquotedblright\ based on whether the manager will receive the investment or not if he reports his type truthfully:

\begin{itemize}
\item Low-type managers with $\theta \leq \theta_{1}^{f}$ belong to the \textquotedblleft no-investment\textquotedblright\ zone in which the project will be scrapped regardless of the signal, and thus receive no rent, i.e., $R(\theta)=0$.

\item High-type managers with $\theta >\theta_{0}^{f}$ belong to the \textquotedblleft unconditional investment\textquotedblright\ zone, in which investment will be made regardless of the signal. The rent for managers in this zone, denoted by $R_{u}(\theta)$, is 
\begin{equation} \label{eq:Ru}
R_{u}(\theta )=\theta w-c.
\end{equation}

\item Intermediate-type managers with $\theta_{1}^{f}<\theta \leq \theta_{0}^{f}$ belong to the \textquotedblleft conditional investment\textquotedblright\ zone, in which investment will be made if and only if the firm receives a good signal. The rent for managers in this zone, denoted by $R_{c}(\theta )$, is 
\begin{equation} \label{eq:Rc}
R_{c}(\theta )=\theta (1-q_1)(w-c)-(1-\theta )q_0c.
\end{equation}
\end{itemize}

\begin{figure}[h]
\begin{center}
\includegraphics[width=1\textwidth]{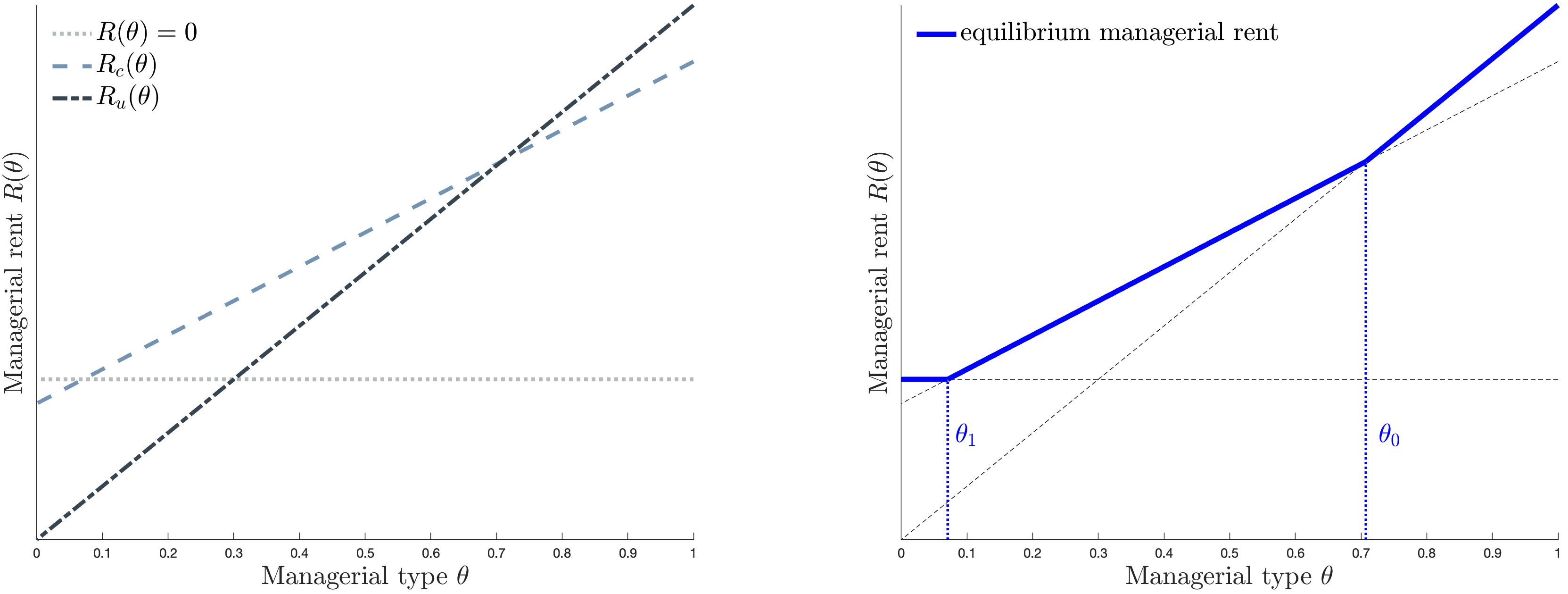}
\end{center}
\caption{{\footnotesize This figure illustrates the construction of the equilibrium managerial rent function $R(\theta)$. The left panel plots the managerial rent for the three investment zones for all $\theta\in(0,1)$: $R(\theta)=0$ (no-investment, dotted line), $R_c(\theta)$ (conditional investment, equation (\ref{eq:Rc}), dashed line), and $R_u(\theta)$ (unconditional investment, equation (\ref{eq:Ru}), dashed-dotted line). The right panel plots the equilibrium rent (blue solid line), which is the upper segment of the individual rent functions shown in the left panel. $\theta_1$ and $\theta_0$ correspond to the left and right boundaries of the conditional investment zone in the equilibrium defined in (\ref{eq:IC_1}) and (\ref{eq:IC_0}), respectively.}}
\label{fig:R1}
\end{figure}

The left panel of Figure \ref{fig:R1} illustrates these three functions: $R= 0$, $R_u(\theta)$, and $R_c(\theta)$ for the entire space of $\theta\in(0,1)$. Note that both $R_u$ and $R_c$ can be lower than zero if $\theta$ is sufficiently low. This is because investment is risky not only to the firm but also to the manager: he has to bear a cost $c$ and only receives the wage $w$ if the project succeeds. This leads to an important observation regarding incentive compatibility: once placed in each zone, the manager's rent is a function of his true type only. Thus, the only potential benefit from misreporting his type is to distort the firm's investment decision by manipulating the firm's belief about the manager's success likelihood. Since the manager can freely choose which investment zone he wishes to be in through his report, a contract is incentive compatible if and only if the manager's equilibrium rent is the upper segments of all three functions ($R = 0$, $R_u$, and $R_c$), as shown in the right panel of Figure \ref{fig:R1}.

The equilibrium managerial rent has two kink points. The left kink, denoted by $\theta_1^m$, represents the manager that receives equal rents from being in the no-investment zone or in the conditional investment zone, and solves 
\begin{align}  \label{eq:rent continuous1}
\theta_1^m(1-q_1)(w-c) - (1-\theta_1^m)q_0c= 0,
\end{align}
where the superscript $m$ indicates that this is a cutoff based on the manager's reporting strategy. The right kink, denoted by $\theta_0^m$, represents the manager that receives equal rents from being in the conditional investment or the unconditional investment zones, and solves 
\begin{align}  \label{eq:rent continuous0}
\theta_0^m(1-q_1)(w-c) - (1-\theta_0^m)q_0c= \theta_0^mw - c.
\end{align}
Incentive compatibility thus requires that all managers form the correct rational expectations regarding the boundaries of each investment zone, and no manager is strictly better off relocating himself into a different zone by misreporting his type. That is, a contract is incentive-compatible only if 
\begin{align}
\theta_1^f & = \theta_1^m \equiv \theta_1,  \label{eq:IC_1}
\\
\theta_0^f & = \theta_0^m \equiv \theta_0.  \label{eq:IC_0}
\end{align}
If one of these two conditions is not met, e.g., if $\theta_0^f <\theta_0^m$, then managers with type $\theta\in\left(\theta_0^f ,\theta_0^m\right)$ will receive unconditional investment if they report their types truthfully. However, their rents are higher if they are in the conditional investment zone, which they can achieve by under-reporting their type, violating incentive compatibility. Intuitively, these are managers with intermediate levels of $\theta$, which represent a modest unconditional probability of success. They are better off being in the conditional investment zone, where they can receive higher rents by free-riding the firm's signal regarding their likelihood of success. Similarly, if $\theta_0^m < \theta_0^f$, some managers with type $\theta\in\left(\theta_0^m,\theta_0^f\right)$ are better off over-reporting their type in order to receive unconditional investment. These are managers with sufficiently high unconditional success probability and can obtain more rents from maximizing the likelihood of receiving the wage payment, which is possible only if the investment is made.

Finally, we prove that although wage is allowed to be made contingent on the manager's reported type, the firm will optimally choose not to do so and instead offer a flat wage $w^*$ to all managers. The complete argument is in the Appendix (Part II of the proof for Proposition \ref{prop:rent}), and the intuition is as follows. First, the optimal wage must be independent of the managers' types within each investment zone; otherwise, all managers in that zone will always report their types to be the one that earns the highest wage. Consequently, there can be at most two levels of wage, one for intermediate-type managers in the conditional investment zone, and one for high-type managers in the unconditional investment zone. We show that if the former is strictly smaller than the latter, the contract is not incentive compatible; if the former is strictly larger than the latter, the contract is incentive compatible but not optimal: the firm can do better by lowering the higher wage for the intermediate-type managers in the conditional investment zone, which improves both the screening effect (i.e., higher $\theta_1$, discussed in details in Section \ref{subsection:screening}) and the firm's profit when the project does succeed (via lowering the wage payment upon success). Therefore, the wage must be a constant under the optimal contract.

Collecting all the results, we now formally summarize the optimal screening contract and the implied equilibrium managerial rent as follows:

\begin{proposition}
\label{prop:rent} Given the firm's information quality (noise) $\mathbf{q} \equiv [q_1\; q_0]^\top$, the optimal incentive compatible screening contract that satisfies (\ref{eq:IC_1}) and (\ref{eq:IC_0}) has the following properties:

\begin{itemize}
\item[i)] The wage for success is a constant, i.e., $w(\theta) = w^*$ for all $\theta$, where
\begin{align}
w^* \equiv \frac{c}{k + c}.  \label{eq:wage}
\end{align}

\item[ii)] The implied boundaries of the investment zones are: 
\begin{align}
\theta_1(\mathbf{q}) & = \frac{q_0(k+c)}{(1-q_1)(1-k-c)+q_0(k+c)},  
\label{eq:tilde theta1}\\
\theta_0(\mathbf{q}) & = \frac{(1-q_0)(k+c)}{q_1 + (1-q_1-q_0)(k+c)}.
\label{eq:tilde theta0}
\end{align}

\item[iii)] The equilibrium managerial rent $R(\theta )$ is given by 
\begin{equation*}
R(\theta )=%
\begin{cases}
0\;,\text{ if }\theta \in (0,\theta_{1}], \\ 
\theta (1-q_1)(w^{\ast }-c)-(1-\theta )q_0c,\text{ if }\theta \in (\theta_{1},\theta_{0}], \\ 
\theta w^{\ast }-c,\text{ if }\theta \in (\theta_{0},1).
\end{cases}%
\end{equation*}
\end{itemize}
\end{proposition}

\subsection{The Screening Effect of Information Quality}

\label{subsection:screening}

Proposition \ref{prop:rent} leads to our main result of this section: how the managerial pool varies with the firm's information quality. Define
\begin{align*}
\nabla X(\mathbf{q}) \equiv 
\begin{pmatrix}
\frac{\partial X}{\partial q_1} \\
\rule{0pt}{1.5em} \frac{\partial X}{\partial q_0}
\end{pmatrix},
\end{align*}
as the \textit{gradient} of a generic function $X$ (i.e., the partial derivatives with respect to $q_1$ and $q_0$), we can formally summarize the following observation:

\begin{proposition}
\label{prop:screening} Under the optimal screening contract characterized in Proposition \ref{prop:rent}, the firm's managerial pool consists of all the managers with types higher than a cutoff level: i.e., $\Theta = \{\theta:\theta>\theta_1(\mathbf{q})\}$, where $\theta_1(\mathbf{q})$ is given by (\ref{eq:tilde theta1}). The cutoff is increasing in the noisiness of the firm's signal, i.e., $\nabla \theta_1(\mathbf{q}) >0$ for all $q_1,q_0$.
\end{proposition}

Proposition \ref{prop:screening} follows naturally from the properties of the optimal screening contract summarized in Proposition \ref{prop:rent}. Because only managers with strictly positive rent in the equilibrium will join the managerial pool, the pool consists of all managers with types higher than $\theta_1(\mathbf{q})$, i.e., all managers in the conditional and unconditional investment zones, and the minimal type of the managerial pool is also the left boundary of the conditional investment zone.

\begin{figure}[h]
\begin{center}
\includegraphics[width=1\textwidth]{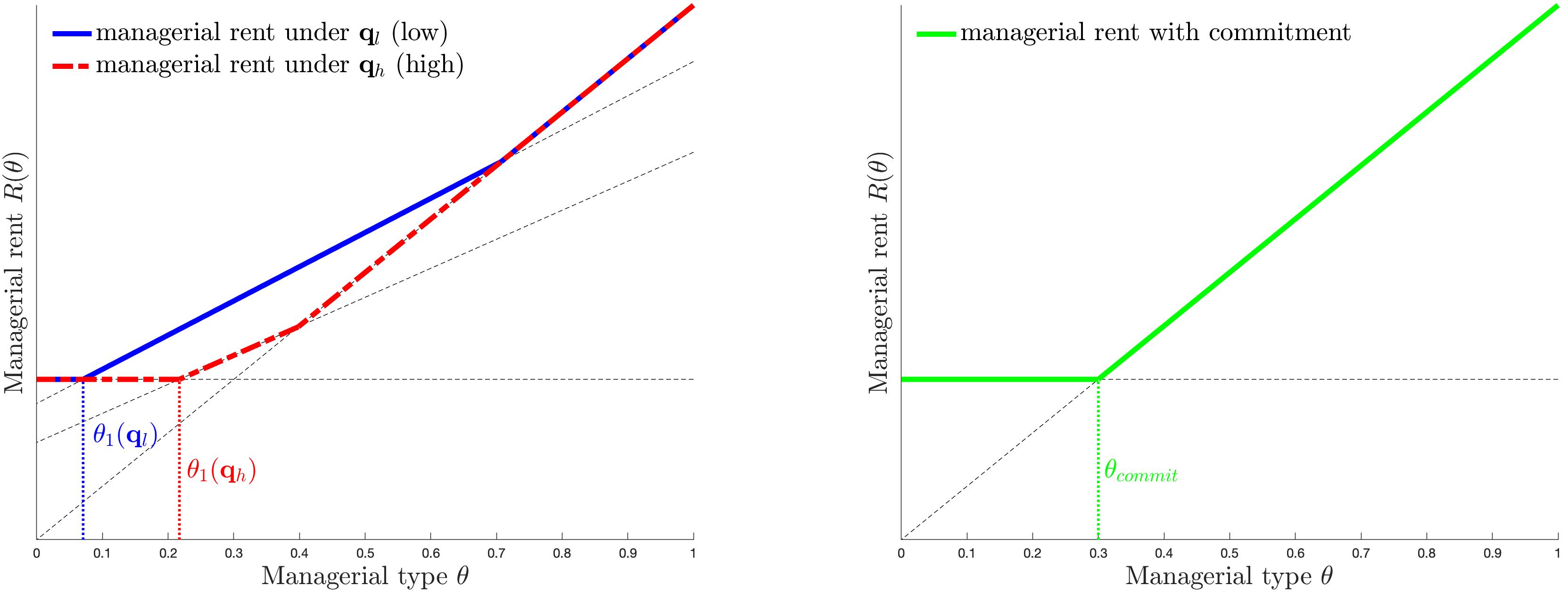}
\end{center}
\caption{{\footnotesize The left panel plots the equilibrium rent functions and the resulting $\theta_1(\mathbf{q})$, the minimal type in the firm's managerial pool, under different levels of information quality $\mathbf{q}_l$ and $\mathbf{q}_h$, where $\mathbf{q}_l < \mathbf{q}_h$. For better illustration, $q_1$ is set to equal $q_0$. The right panel plots the rent function and the resulting $\theta_{commit}$, the minimal type in the firm's managerial pool, for the ``with-commitment" benchmark discussed in Section \ref{subsection:role of commitment}.}}
\label{fig:R2}
\end{figure}

Importantly, $\theta_{1}(\mathbf{q})$ is an \textit{increasing} function in both $q_1, q_0$, meaning that a noisier signal achieves more effective screening of managers by excluding more low-type candidates from the managerial pool. We refer to this as the (positive) \textit{screening effect} of noisier information. Figure \ref{fig:R2} (left panel) visualizes such screening effect by plotting two rent functions under different levels of information quality and their corresponding $\theta_{1}$. Intuitively, in the equilibrium, $\theta_{1}$ solves 
\begin{equation}
p_{1}(\theta_{1})=\frac{k}{1-w^{\ast }}=k+c,
\end{equation}%
where $p_{1}(\theta)$, given in (\ref{eq:p1}), is an increasing function of $\theta$ and a decreasing function of both $q_1$ and $q_0$. The more precise the signal, the higher $p_{1}(\theta)$ is for any $\theta$, once a good signal is observed. Consequently, higher-quality information (i.e. lower $\mathbf{q}$) lowers the minimal type of manager with whom the firm is willing to invest upon observing a good signal. This expands the left boundary of the conditional investment zone, allowing more managers with lower $\theta $ to have the opportunity to trigger investment. These managers, in turn, can expect to receive positive rents and thus are willing to join the managerial pool. 

In contrast, when information quality decreases, $p_{1}(\theta)$ is less informative about the likelihood of success. While an increase in either $q_1$ or $q_0$ achieves qualitatively the same result, the two sources of noise manifest via different channels. An increase in $q_1$ lowers the precision of the good signal as an indicator of success, leading to more ``false negatives" in the investment decision based on the signal. Conversely, an increase in $q_0$ increases the likelihood that the good signal is an indicator of failure, thus leading to more ``false positives" in the investment decision based on the signal. Nevertheless, both sources of noisiness reduce the accuracy of inferring the underlying state from the signals and thus the firm's reliance on the signals for investment decisions. As a result, only managers with sufficiently high types can trigger the investment even under a good signal, and more low-type managers are left in the no-investment zone and thus ``screened'' out of the managerial pool.

In sum, the firm in our model has two objectives: screening for high-ability managers, and investing only when it knows the project will succeed. The benefit of a better signal is reflected in the latter objective by improving the firm's assessment of the likelihood of success from the unconditional probability $\theta$ to the conditional probability $p_1(\theta)$. However, this improvement of the success likelihood also benefits the intermediate-type managers with modest unconditional success probability, who otherwise would not be allowed to invest without a (good) signal. From these managers' perspective, they bear the personal cost $c$ if the investment is made and only receive payments when the investment succeeds. Thus, managers are most concerned about the firm making a type-I error (i.e., investing without success) and derive an \textit{option value} from the firm's information, from which the firm can gain additional confidence in the likelihood of success before investing. Because a more precise signal reduces the likelihood of such type-I investment errors, it improves the option value and thus lowers the minimal type of managers willing to accept the firm's contract, undermining the firm's first objective of screening. In comparison, lower information quality reduces the accuracy of the firm's assessment of the success likelihood as well as the option value of the signal. This hurts not only the firm but also the intermediate-type managers, thus allowing the firm to achieve more effective screening by excluding more of those managers from the managerial pool.

It is also noteworthy that, when studying the firm's investment and hiring decisions, we do not exogenously bundle the two decisions.\footnote{In fact, our paper models the two decisions separately, capturing the firm's investment decision as one contingent on the noisy signal $s$ about the investment outcome, and the hiring/screening decision as designing a screening contract (the wage for a successful investment) to illicit truthful reporting from the manager, and making the hiring decision based on the manager's reported type $\hat\theta$.} Instead, the two decisions are connected not because they are based on the same information signals. The signal $s$ is used only in the investment decision but \textit{not} directly in the hiring decision since $s$ is produced after the manager is hired. Rather, we show that the firm's investment and hiring decisions are endogenously intertwined through subtle equilibrium effects of varying the quality of $s$, as discussed above. 

\subsection{The Role of Commitment} \label{subsection:role of commitment} 

We have so far established how the screening of managers depends on the quality of the firm's information. Given this is the central result of the paper, we offer in this section an alternative understanding of its intuition based on its connection with the firm's (lack of) commitment power. Readers who are more interested in the implications of our model can skip this section and move on to the next.

Consider the case in which the firm \textit{can ex ante} commit to any investment policy regardless of the signal. Suppose the firm's objective is to achieve the best screening result (i.e., maximizing the minimal type in the managerial pool) while still offering the equilibrium wage $w^* = c/(k+c)$ found in (\ref{eq:wage}) for success.\footnote{Technically speaking, with commitment, the firm can offer any wage $w>c$ because the IC conditions (\ref{eq:IC_1}) and (\ref{eq:IC_0}) are no longer relevant when the firm can commit to any $\theta^f_s$. We assume the firm still offers the equilibrium wage $w^*$ here to simplify the illustration of the role of commitment.} The firm's optimal policy is to offer $w^*$ to all managers and commit to scrapping (continuing) all investments if the manager's reported type is less (more) than $\theta_{commit} =c/w^*$. In other words, the space of managers is partitioned into two zones only: the ``no investment'' zone for low-type managers (i.e., $\theta <\theta_{commit}$) who also earn zero rent, and the ``unconditional investment'' zone for high-type managers (i.e., $\theta > \theta_{commit}$) who earn positive rents. The right panel of Figure \ref{fig:R2} illustrates the managerial rent and the resulting managerial pool under this contract. The firm achieves the best possible screening result because $\theta_{commit}=c/w^{\ast }>\theta_{1}(q_1, q_0)$ (given in equation (\ref{eq:tilde theta1})) for any $\{q_1, q_0\}$. Intuitively, the ability to commit to investment policies eliminates the ``conditional investment" zone, where the firm's private signal is used. This lowers the rent for intermediate-type managers (to zero) and thus excludes them from the managerial pool.

Without commitment, the firm is unable to sustain this screening result. The lack of commitment creates a conditional investment zone for intermediate-type managers who rationally expect some rents in the equilibrium. This is because $p_1(\theta)$, the conditional probability of success, is sufficiently high for these managers. Consequently, while the firm would like to exclude most of them from the managerial pool by committing never to invest, it is not sub-game perfect to do so once a good signal is observed. A more precise signal (i.e., lower $q_1$ or $q_0$) increases $p_1(\theta)$ for all $\theta$, thus lowering the minimal type of managers that expects to receive investment (and thus positive rents) in equilibrium, drawing them to the managerial pool.\footnote{In addition, the lack of commitment implies that the firm cannot offer any wage $w<w^*$ while maintaining incentive compatibility. It can be shown that $\theta_0^f<\theta_0^m$ if $w<w^*$. Therefore, managers with type $\theta\in(\theta_0^f, \theta_0^m)$ will have the incentive to misreport their type and be placed into the conditional investment zone, which yields them higher rents than reporting truthfully and being placed into the unconditional investment zone. Therefore, compared to the ``with-commitment" benchmark, the lack of commitment raises the wage needed to maintain incentive compatibility and creates the conditional investment zone that attracts some intermediate-type managers to the managerial pool.}

In contrast, a noisier signal improves the firm's screening result by lowering the conditional probability of success given the good signal, causing the firm to scrap more often for the intermediate-type managers. This intuition can be best illustrated in the extreme case in which the signals are completely uninformative (i.e., $q_1=q_0=1/2$). This is equivalent to no signal being available, and the firm only makes investment decisions based on the reported managerial type. Consequently, the equilibrium coincides with the \textquotedblleft with-commitment\textquotedblright\ equilibrium described above, where the firm achieves the best screening result.\footnote{Of course, the \textquotedblleft with-commitment\textquotedblright\  equilibrium is not necessarily optimal for the firm overall, due to the same trade-off between better screening and more informed investment described in the previous section.} In short, the effect of the precision of the signal on the firm's screening result is related to the friction generated by the lack of commitment. Noisier information mitigates the firm's lack of commitment issue by lowering the usefulness of the signal, thus increasing the likelihood of the scenarios in which the firm can commit to investing only when the manager's type is sufficiently high.

\subsection{Information Quality and Accounting Conservatism} \label{section:conservatism}

For the purpose of generality, we allow the quality of the two signals $\{q_1, q_0\}$ to be independent and potentially asymmetric (i.e., $q_1\neq q_0$). It is, therefore, interesting to explore the underlying mechanisms that drive such asymmetry and their impact on screening. One particular mechanism responsible for asymmetric information content that has received extensive attention in the literature is accounting conservatism, a downward bias such that bad signals are more likely to be disclosed  than good signals. In prior models of accounting conservatism (e.g., \citealp*{gigler2009accounting}; \citealp{caskey2017corporate}, etc.), different degrees of conservatism result in different levels of noisiness in the good versus the bad signals. In our model, we can likewise separate the firm's degree of accounting conservatism, which we denote by $\lambda$, from the (symmetric) noisiness of the signals, which we denote by $q$, as follows:
\begin{align}
\Pr(s = 0|\pi = 1) = q + \lambda, \\
\Pr(s = 1|\pi = 0) = q - \lambda.
\end{align}
That is, while information noise $q$ affects the likelihood that the signal is the opposite of the underlying state, conservatism $\lambda$ always increases the likelihood of a bad signal. We assume $\lambda<q<1/2$ so that both probabilities above are well defined. Put differently, given $\theta$, the probability of all events can be summarized as follows:
\vspace{0.1in}
\begin{center}
\begin{tabular}{rcc}
\hline\hline
& Success $\pi= 1$ & Failure $\pi= 0$ \\ \hline
Good signal $s= 1$ & $\theta (1-q-\lambda)$ & $(1-\theta)(q-\lambda)$ \\ 
Bad signal $s= 0$ & $\theta (q + \lambda)$ & $(1-\theta)(1-q+\lambda)$ \\ 
\hline\hline
\end{tabular}
\end{center}
\vspace{0.1in}
Note that this information structure is equivalent to that of our baseline model with the following change of variables:
\begin{align}
q & = \frac{q_1 + q_0}{2}, \\
\lambda & = \frac{q_1 - q_0}{2}.
\end{align}
In other words, when $q_1\geq q_0$, they can be decomposed into the component that corresponds to the noisiness of \textit{both} signals, and the component that increases the noisiness of the bad signal only. Since this decomposition does not change the underlying model, the same solution characterized in Proposition \ref{prop:rent} remains intact. Using the new variables \{$q, \lambda$\}, the boundaries of the conditional investment zones (equations (\ref{eq:tilde theta1}) and (\ref{eq:tilde theta0})) can be re-written as:
\begin{align}
\theta_1(q,\lambda) & = \frac{(q-\lambda)(k+c)}{(1-q-%
\lambda)(1-k-c)+(q-\lambda)(k+c)}  \label{eq:theta_1_conservatism}, \\
\theta_0(q,\lambda) & = \frac{(1-q+\lambda)(k+c)}{q + \lambda +
(1-2q)(k+c)}.  \label{eq:theta_0_conservatism}
\end{align}
Simple algebra therefore implies the following result:

\begin{proposition} \label{prop:conservatism}
The minimal type in the firm's managerial pool is increasing in the noisiness of the firm's signal while decreasing in the firm's conservatism, i.e., 
\begin{align}
\frac{\partial \theta_{1}(q,\lambda)}{\partial q} > 0, \quad \text{ and } \quad \frac{\partial \theta_{1}(q,\lambda)}{\partial \lambda} < 0,
\end{align}
for all $0 \leq \lambda<q<1/2$.
\end{proposition}

\begin{figure}[h]
\begin{center}
\includegraphics[width=1\textwidth]{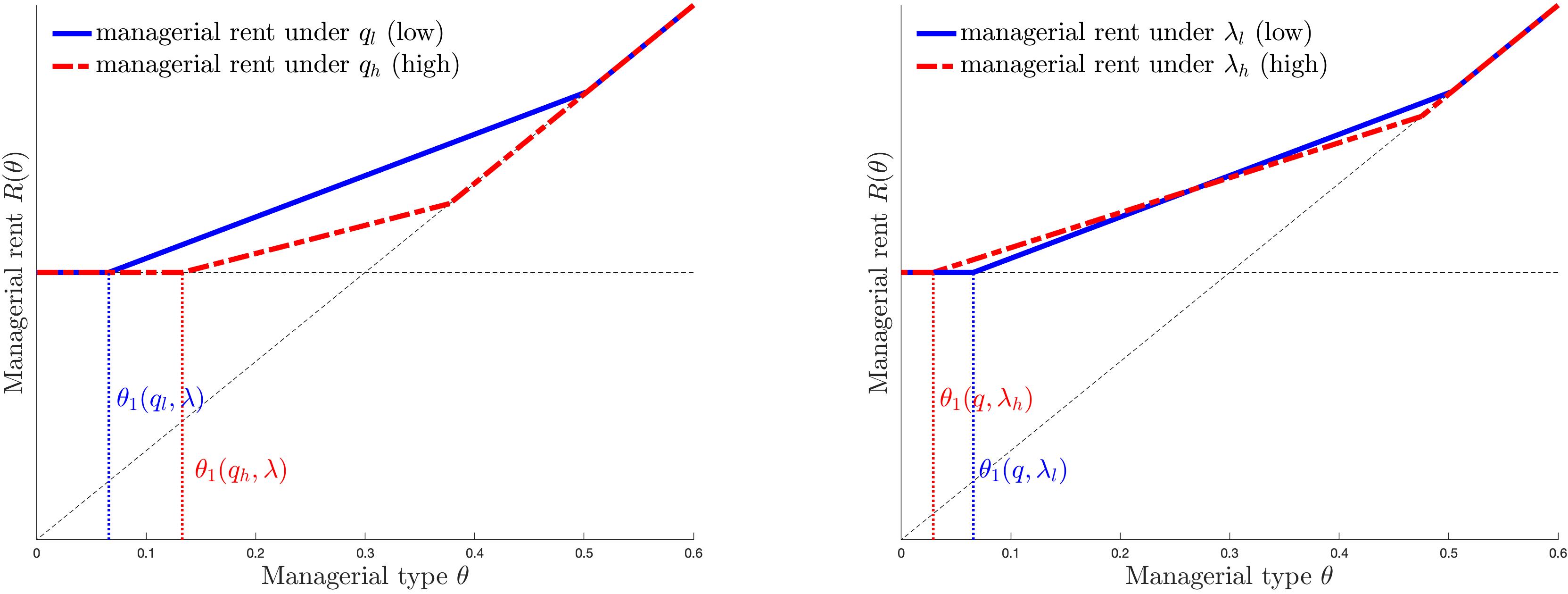}
\end{center}
\caption{{\footnotesize This figure illustrates the different screening effects of information noise $q$ and accounting conservatism $\lambda$. The left panel shows the effect of an increase in $q$ (i.e., $q_h>q_l$) while holding $\lambda$ constant. The right panel shows the effect of an increase in $\lambda$ (i.e., $\lambda_h>\lambda_l$) while holding $q$ constant. In both panels, the blue solid line corresponds to a smaller $q$ or $\lambda$, while the red dashed line corresponds to a larger $q$ or $\lambda$. $\theta_1$ corresponds to the minimal type in the managerial pool as given in (\ref{eq:theta_1_conservatism}).}}
\label{fig:2controls}
\end{figure}

Figure \ref{fig:2controls} illustrates the different screening effects of information noise and conservationism. The effect of the former is depicted in the left panel (which is qualitatively analogous to the left panel of Figure \ref{fig:R2}). An increase in the information noisiness means the signal is less valuable. The firm thus relies less on the signal for the investment decision, which leads to a smaller conditional investment zone and lower managerial rent in that zone, generating a positive screening effect.

In contrast, the effect of conservatism, depicted in the right panel of Figure \ref{fig:2controls}, suggests the opposite screening effect. This is due to two distinct effects of conservatism on the information content of each signal.\footnote{These properties are also highlighted in \citet{caskey2017corporate} (Assumptions A2 and A3).} On the one hand, more conservative accounting reduces the information content of the bad signal, thus increasing the conditional probability of success when the firm receives a bad signal. As a result, the firm is less willing to scrap the investment for lower-type managers, which means that $\theta_0$, the right boundary of the conditional investment zone, decreases as $\lambda$ increases. However, when the firm is more conservative, the good signal becomes \textit{more informative} of actual success. As a result, the firm is also more willing to continue investment conditional on a good signal, even for a low-type manager. This expands the left boundary of the conditional investment region as $\theta_1$ decreases. This effect is particularly strong for the low-type managers who are far away from the right boundary of the conditional investment region (where an increase in type leads to investment even under the bad signal). Altogether, the conditional investment region shifts more to the left under more conservative accounting systems. Since the screening effect is determined by the left boundary of the conditional investment zone, more conservatism leads to a decline in $\theta_1$.

In short, while both information noisiness and accounting conservatism affect the degree to which the firm relies on the signal for the investment decisions, the two have contrasting impacts on the screening of managers. Noisier information leads to positive screening (i.e., an improvement of the minimal type in the managerial pool), while more conservative accounting leads to the opposite screening result. These results contribute to the literature on the role of accounting conservatism in firms' investment decisions, such as \citet{nan2014financing}, \citet{laux2020effects} \citet{laux2021accounting}, etc. Notably, \citet{caskey2017corporate} find that conservative accounting improves the oversight of managers' investment decisions, but at the same time, provides managers with incentives to distort the accounting system. \citet{bagnoli2005conservative} examine a firm's choice of conservative reporting policies, and derive the conditions under which such choice serves as a signal for the investors to infer the firm's private information. \citet*{dordzhieva2022signaling} show that firms may choose to adopt a liberal accounting system in order to signal favorable private information and attract external financing. In this literature, the optimality of conservative or liberal systems often results from the trade-off between the costs of false negative and false positive errors. In our model, the costs and benefits of noisy or biased information systems are not driven by these two errors. Instead, the trade-off emerges from the \textit{ex ante} more effective screening of managers and the \textit{ex post} more informed investment decisions.

\section{Information Quality and Firm Value}  \label{section:firm value} 

We are now ready to examine the overall relationship between information quality and the \textit{ex ante} firm value $V(\mathbf{q})$, defined in (\ref{eq:V(q) definition}). Given the contract described in Proposition \ref{prop:rent}, $V(q)$ can be written as 
\begin{align}
V(\mathbf{q}) & = \left(\frac{1}{1-F(\theta_1)} \right)\left[\int^{\theta_0 }_{\theta_1}v_c(\theta)dF(\theta) +\int^{1}_{\theta_0}v_u(\theta)dF(\theta) \right], \label{eq:V(q)_equilibrium}
\end{align}
where
\begin{align}
v_c(\theta) & = \theta(1-q_1)(1-w^* - k) - (1-\theta)q_0k, \label{eq:v_c}\\
v_u(\theta) & = \theta(1-w^*) - k, \label{eq:v_u}
\end{align}
and $w^*$, $\theta_1$, and $\theta_0$ are given in (\ref{eq:wage}), (\ref{eq:tilde theta1}), and (\ref{eq:tilde theta0}). $v_c(\theta)$ and $v_u(\theta)$ represent respectively the firm's expected payoff from hiring a $\theta$-type manager in the conditional and unconditional investment zones.

We provide three sets of results regarding firm value $V(\mathbf{q})$. Section \ref{subsection:firm value_marginal} derives the marginal value of information quality. Section \ref{subsection:firm value_end points} explores and compares some information design choices. Finally, Section \ref{subsection:comp stats} discusses several comparative statics. 

\subsection{Marginal Value of Information Quality} \label{subsection:firm value_marginal} 

We first analyze the marginal impact of information quality on overall firm value, i.e., $\nabla V(\mathbf{q})$, which manifests through three channels: $\theta_1$, the lowest type in the managerial pool; $\theta_0$, the boundary between the conditional and unconditional investment zones; and the slope of $v_c(\theta)$ in the conditional investment zone. The first channel corresponds to the screening effect: the firm's objective of hiring the best manager. The second and third channels reflect the firm's objective of investing when the project is likely going to succeed. To better illustrate the impact of information quality on firm value through these two objectives, differentiating (\ref{eq:V(q)_equilibrium}) with respect to $\mathbf{q}$ and observing that $v_c(\theta_1)=0$ and $v_c(\theta_0)=v_u(\theta_0)$ in the equilibrium, we have:
\begin{align}
\nabla V(\mathbf{q})  & = \frac{\partial V}{\partial \theta_1}\nabla \theta_1(\mathbf{q}) + \frac{\partial V}{\partial \theta_0}\nabla\theta_0(\mathbf{q})  + \left(\frac{1}{1-F(\theta_1)} \right)\left[\int^{\theta_0 }_{\theta_1}\nabla v_c(\mathbf{q}) dF(\theta)\right] 
\\
& = \underbrace{V(\mathbf{q})\left(\frac{f(\theta_1)}{1-F( \theta_1)} \right)\nabla\theta_1(\mathbf{q})}_{\text{screening effect, } >0 } + \underbrace{\left(\frac{1}{1-F(\theta_1)} \right)\left[\int^{\theta_0 }_{\theta_1}\nabla v_c(\mathbf{q})dF(\theta)\right]}_{\text{change in investment profit, } <0}.  \label{eq:V(q)_decompose}
\end{align}
The first term in (\ref{eq:V(q)_decompose}) reflects the marginal screening effect, which is positive because $\nabla\theta_1(\mathbf{q})>0$ (Proposition \ref{prop:screening}). The second term in (\ref{eq:V(q)_decompose}) reflects the marginal change in the firm's investment profit in the conditional investment zone, which is negative because $\nabla v_c(\mathbf{q})<0$. That is, a noisier signal erodes the firm's expected profit in the conditional investment zone by reducing the informativeness of the good signal. Altogether, (\ref{eq:V(q)_decompose}) demonstrates the trade-off the firm faces when designing an optimal information system.

\begin{figure}[h]
\begin{center}
\begin{tabular}{cc}
\includegraphics[width=0.45\textwidth]{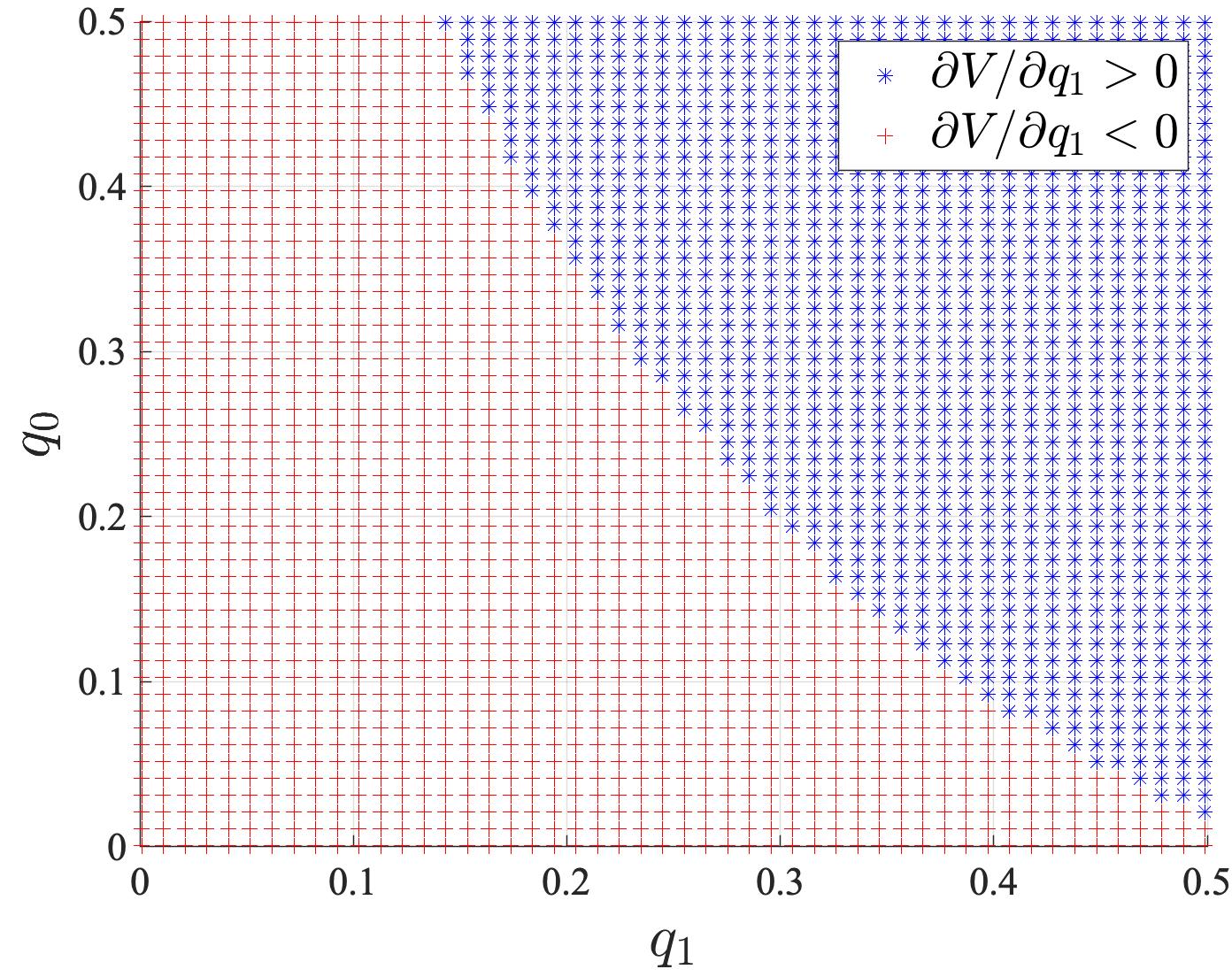} & 
\includegraphics[width=0.45\textwidth]{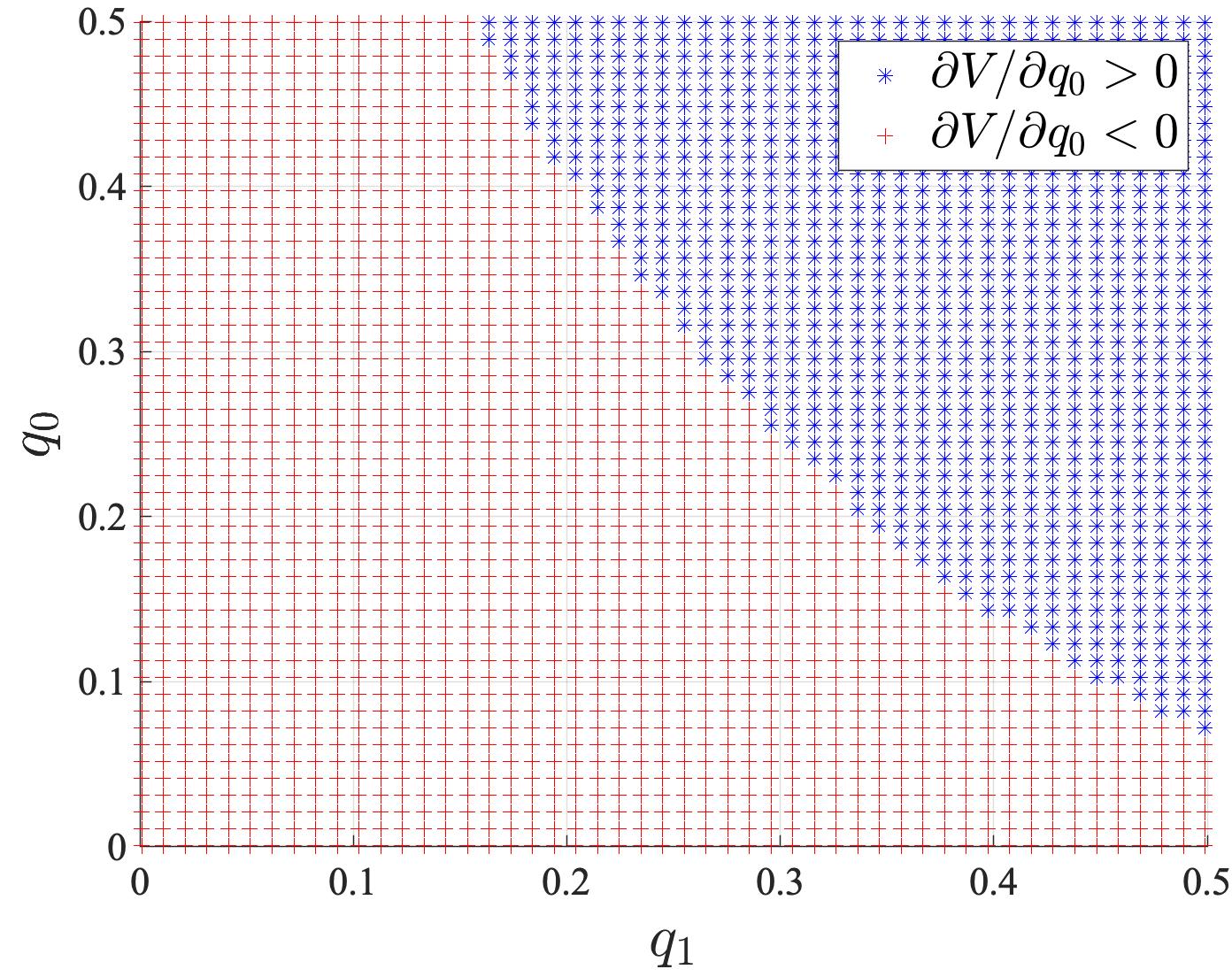}
\end{tabular}
\end{center}
\caption{{\footnotesize This figure plots the sign of $\partial V/\partial q_1$ (left panel) and the sign of $\partial V/\partial q_0$ (right panel) for different values of $q_1,q_0$. A blue asterisk indicates a positive derivative (i.e., positive marginal value) and a red plus indicates a negative derivative (i.e., negative marginal value). $c =k=0.24$, and $\theta$ follows a uniform distribution.}}
\label{fig:V}
\end{figure}

Figure \ref{fig:V} depicts a numerical example of $\nabla V(\mathbf{q})$. Because $\mathbf{q}$ is two-dimensional, for better visualization, these plots focus on displaying the \textit{signs} of $\nabla V(\mathbf{q})$ rather than their exact values. That is, the bottom left corner of each picture shows the sign of either $\partial V/\partial q_1$ or $\partial V/\partial q_0$ when $q_1=q_0\rightarrow 0$, and the top right corner of each picture shows the sign of either $\partial V/\partial q_1$ or $\partial V/\partial q_0$ when $q_1=q_0= 1/2$. Notably, $\nabla V(\mathbf{q})$ changes signs from negative to positive when information becomes noisier, implying a non-monotonic relationship between information quality and firm value. While proving the sign of the marginal impact for arbitrary levels of information quality turns out to be analytically challenging, we can nonetheless prove the key takeaway---that neither more nor less precise information is always necessarily better for the firm---under certain conditions:

\begin{proposition} \label{prop:Vprime}
If $\theta$ follows a uniform distribution, then there exists $0<\underline q<\overline q<1/2$ such that $\nabla V(\mathbf{q}) < 0$ if $q_1 <\underline q$ or $q_0<\underline q$, and $\nabla V(\mathbf{q}) > 0$ if $q_1 >\overline q$ and $q_0 > \overline q$. 
\end{proposition}

Proposition \ref{prop:Vprime} states that the screening effect dominates when the firm's information quality is sufficiently low, and an improvement of  information quality (i.e., lowering $\mathbf{q}$) for these firms may actually decrease their value. This can be easily seen algebraically in (\ref{eq:V(q)_decompose}): when $q_1=q_0=1/2$ (meaning that both signals are useless), $\theta_0=\theta_1$, and only the first term that corresponds to the screening effect exists. By continuity, $\nabla V(\mathbf{q})>0$ when both signals contain sufficiently large noises. Intuitively, the benefit from the signal is more informed investment in the conditional investment zone. When the signal is not very informative, the firm relies less on the signal for investment. The conditional investment zone shrinks, manifesting mostly the screening effect. In contrast, when either signal is very precise (i.e., either $q_1$ or $q_0$ is close to zero), the firm makes most of its investment decisions based on the signal. The conditional investment zone is wide, and any deterioration of the precision of the signal has a substantial negative effect on the return to investment in that zone and thus, the overall firm value (i.e., $\nabla V(\mathbf{q})<0$).\footnote{These results are consistent with those in \citet{goldstein2019good} and \citet{meng2020board}, where the marginal value of better (worse) information is positive when information quality is sufficiently high (low).}

These results have useful practical implications. First, they demonstrate that \textit{marginal} improvements in information quality are not always beneficial. In practice, producing useful information for operational decisions such as investment is often costly and requires a substantial amount of subjective judgment, rendering a perfect information system infeasible and any significant improvement of an existing system challenging. Given these constraints, the overall impact of gradual adjustments to the information system should be evaluated with caution. Second, in our main analysis, we have interpreted the signal $s$ as the firm's preliminary assessment of the project's prospect, and its quality stems from the characteristics of the firm's board members making such assessment, such as their background, experience, and industry expertise. It is natural to expect local and gradual variations of these firm characteristics, especially in cross-sectional/time-series empirical analyses. Accordingly, one may draw implications from our analysis of the marginal effect of increasing the information quality to guide empirical tests of the relationship between firms' information environments and firm value. Lastly, our results also provide a possible explanation for the observed heterogeneity in both the quality of internal information systems and how information is utilized. In practice, firms vary substantially in their manners and intensities of conducting internal monitoring or auditing, and managers are allowed various degrees of latitude in making operational decisions. Our results suggest that such diversity could be the result of firms optimally balancing the direct effect of more informed decisions and the indirect screening effect of information. In particular, a lack of frequent monitoring and auditing or a large degree of liberty for the managers is not necessarily the result of failure in internal management. 

\subsection{Information Design} \label{subsection:firm value_end points} 

So far we have focused on characterizing the effects of varying the quality of the firm's information system, treating such information quality as exogenously imposed. Thus, the marginal value in information quality can be interpreted as the result of exogenous variations in firms' information environments. Nonetheless, it is also interesting to explore how the firm would design its information system \textit{ex ante} if it can choose the information quality. We consider such analysis in this section. 

Specifically, to study the firm's information design choice, we make the following two modifications to our main model. First, at $t =0$ before offering the compensation contract, the firm can \textit{ex ante} choose the quality of its information system, i.e., the precision of the signal that will be produced at $t=1$. For simplicity and tractability, we focus on the case in which the precision of the signal is symmetric: $q_1=q_0=q$.\footnote{Deriving the firm's optimal information choices of $\{q_1,q_2\}$ when $q_1 \neq q_2$ requires imposing additional structures (for instance, we need to specify how the information production cost of reducing $q_1$ interacts with the cost of reducing $q_2$), which complicates the analysis and obfuscates the insights of the model.} Second, the firm incurs a cost $h(q)$ for producing the signal $s$ with quality $q$, where $h'(q) < 0$ and $h''(q) > 0$. That is, more precise information is more costly to obtain, and the marginal cost for improving the information quality (i.e., reducing information noise $q$) is increasing in the quality of information.\footnote{In our main model, we assume that it is costless to produce the signal $s$ to focus on the endogenous benefit and cost of more precise information in improving the investment efficiency and in distorting the screening of managers, respectively. These endogenous benefits/costs, however, do not make the firm value \textit{concave} in information quality, which is necessary for attaining an interior solution. In fact, absent the information production cost, the firm value is U or V-shaped in information quality so that the firm value is maximized at one of the two extremes (e.g., either no information or perfect information). In this light, introducing the information production cost $h(q)$ helps to ensure the concavity of the net firm value $V(q)-h(q)$ in the information quality $q$, which then allows us to characterize the interior choice of $q$ that maximizes such value. Similar cost assumptions have been made in prior studies, e.g., \citet{gao2013informational}, \citet{zhang2021competition}, etc.}

\begin{figure}[h] 
\begin{center}
\begin{minipage}{0.5\textwidth} 
\centering
\includegraphics[width=1\textwidth]{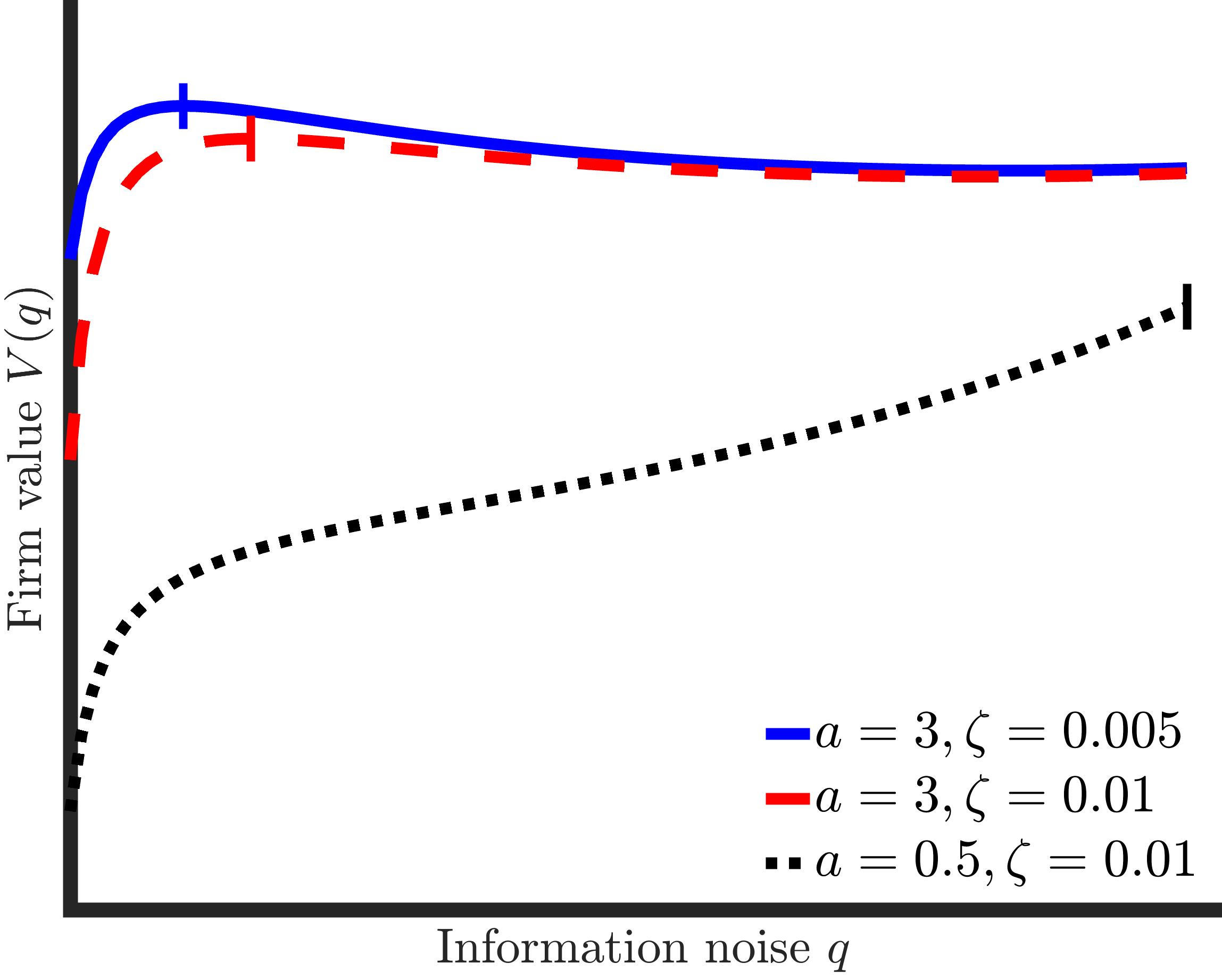} 
\end{minipage}
\hspace{0.05\textwidth} 
\begin{subtable}{0.4\textwidth}
\centering	
\begin{tabular}{c}
\\	
\includegraphics[width=1\textwidth]{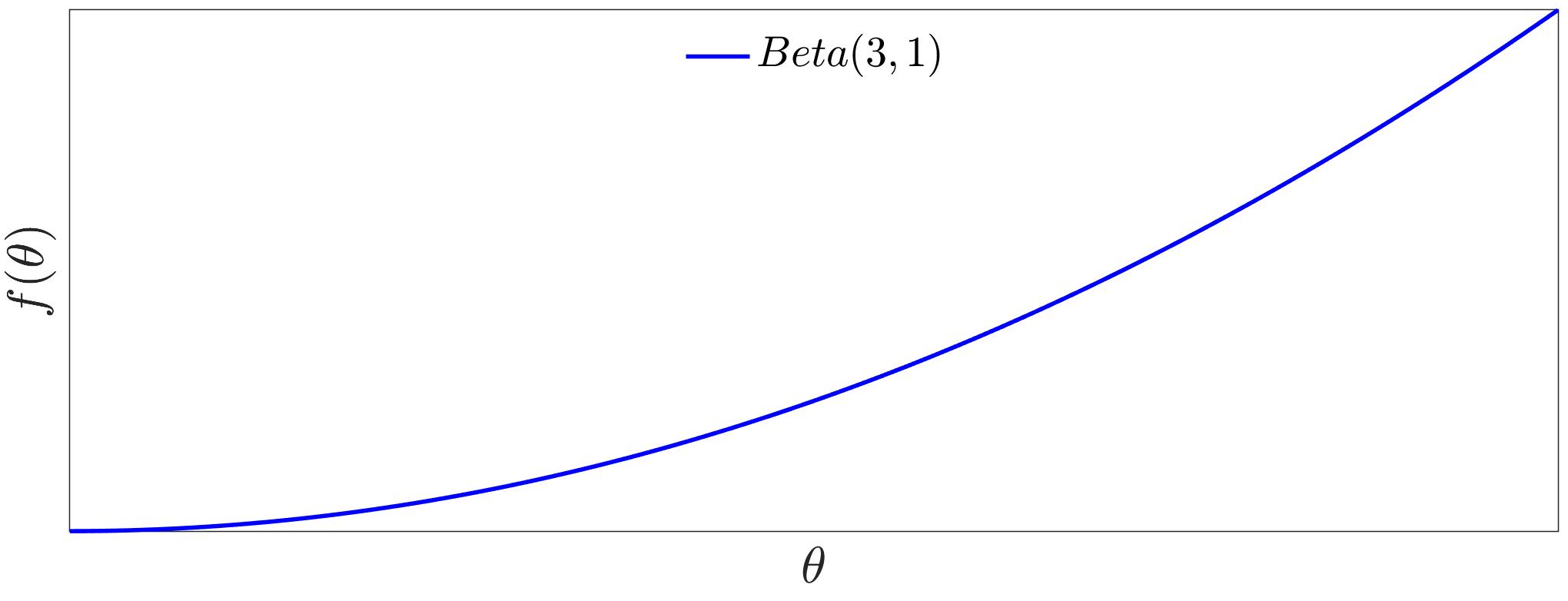} \\
\\
\includegraphics[width=1\textwidth]{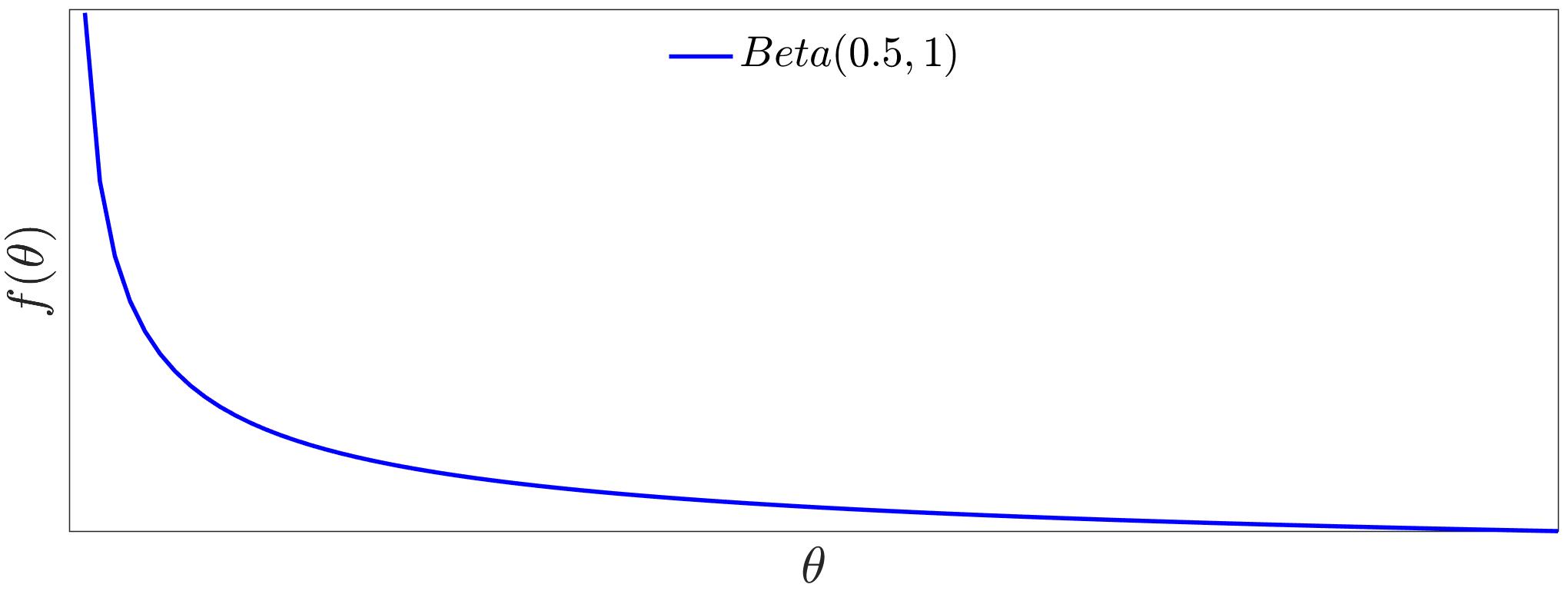} 
\end{tabular}
\end{subtable}
\end{center}
\caption{The left panel of this graph illustrates the shape of the firm's value function and its optimal information choice of $q$. The cost function is $h(q) = \zeta/q$ with $\zeta>0$. The blue solid line corresponds to $\zeta = 0.01$; the red dashed line and the black dotted line correspond to $\zeta = 0.005$. The vertical bars indicate the positions of the optimal information quality $q^*$ in each case: i.e., $q^*\equiv \arg\max_q V(q)-h(q)$.  $\theta$ follows a $Beta(a,b)$ distribution with $a=3$ and $b =1$ for the blue and red lines and $a = 0.5$ and $b = 1$ for the black line. The right panel of the figure depicts the density functions of each distribution of $\theta$ used. $k=c=0.15$ for all cases.} \label{fig:interior}
\end{figure}

With these modifications, the firm's optimal information quality $q^*$ is now defined as:
\begin{align}
q^* = \arg\max_{q\in(0,1/2]}\; V(q) - h(q),
\end{align}
where $V(q)$ is given by the system defined in (\ref{eq:V(q)_equilibrium}) to (\ref{eq:v_u}). Unfortunately, a full analytical characterization
of the optimal information choice is intractable, so we turn to numerical analysis.
Figure \ref{fig:interior} illustrates a few examples of the firm's optimal information choices (marked by the vertical bars on each line). In our numerical analysis, we assume that the information production cost has a functional form $h(q) = \zeta/q$, where $\zeta>0$ captures the marginal cost of obtaining more precision information. It is straightforward to verify that this cost function satisfies all the assumptions we imposed on $h(q)$ earlier. We assume $\theta$ follows a $Beta(a,b)$ distribution, which is theoretically the most flexible distribution on the unit interval. We fix $b = 1$ and vary the parameter $a$ to produce different shapes of distribution for $\theta$, as illustrated by their density functions in the right panel of Figure \ref{fig:interior}.

The examples in Figure \ref{fig:interior} demonstrate the trade-off the firm faces when designing the quality of its information system. On the one hand, better information improves the efficiency of investment once a manager is hired. On the other hand, more precise information is more costly to produce and causes the managerial pool to expand downward (i.e., the screening effect), reducing the expected quality of the manager the firm will be matched with. Interestingly, the trade-off can yield both interior solution for the firm's optimal information quality (the blue solid and red dashed lines in Figure \ref{fig:interior}) as well as the corner solution (the black dotted line in Figure \ref{fig:interior}), where the firm opts for a completely uninformative signal. These differences arise due to the different magnitude of the screening effect, or how many of the low- and intermediate-type managers the firm can exclude from its managerial pool by reducing the precision of its information system. To see this, note that, in our numerical examples, the firm chooses to produce no information when $a<1$. As illustrated in the lower-right panel of Figure \ref{fig:interior}, this represents the case in which the distribution of managerial type has a heavy left tail. Accordingly, the benefit of screening is very strong because it allows the firm to exclude more low-type managers from the managerial pool. The screening effect of coarser information thus dominates, inducing the firm to choose no information production.\footnote{To clarify the screening effect of information when the firm chooses to produce no information, recall that in Proposition \ref{prop:screening}, the screening effect is defined as how a \textit{marginal} change in the information noise $q$ affects the lower-bound $\theta_1$ of the managerial pool. Accordingly, at the corner case of $q = 1/2$, the screening effect corresponds to $\left.\partial \theta_1(q)/\partial q\right|_{q=1/2}$. Such derivative is well-defined at $q = 1/2$, and it captures how $\theta_1(q)$ changes if the firm switches from producing no information ($q=1/2$) to producing a signal with a small amount of precision ($q$ close to $1/2$).} In contrast, when the fraction of low-type managers is smaller (i.e., $a>1$ as in the upper-right panel of Figure \ref{fig:interior}), the screening benefit of noisy information is weaker. Accordingly, the firm chooses an interior level of information quality, balancing the benefit of information in facilitating investments versus its adverse screening effect and the cost of producing information. Intuitively, Figure \ref{fig:interior} shows that when the information production cost $\zeta$ increases, the optimal information quality $q^*$ increases (i.e., noisier information).

Our analysis of firms' information choices can be related to the broad literature on the design of internal information systems. For instance, \citet{bertomeu2015asset} study the optimal asset measurement rules for assets used as collateral. \citet{jiang2017properties} study the accounting rules to reduce firms' signaling cost via excessive share retention. \citet*{friedman2020optimal,friedman2022rationale} consider the \textit{ex ante} value of information systems when managers can acquire and/or disclose additional private information. \citet*{bertomeu2021strategic} study the value of measurement systems when there is uncertainty regarding whether verifiable information exists. It is noteworthy that most of these existing studies have been focusing on the \textit{signaling} role of firms' information design choices.\footnote{In comparison, a few other studies consider the signaling role of firms' \textit{ex post} reporting discretion, e.g., \citet{stocken2004financial}, \citet{fan2007earnings}, \citet{baldenius2010signaling}, etc.} A main departure of our paper from the extant literature is then to identify a \textit{screening} role of the information choice when firms face adverse selection problems in hiring managers. Thus, our results highlight a novel mechanism and complement the literature on the \textit{ex ante} value of firms' information choices.

\subsection{Comparative Statistics} \label{subsection:comp stats}

This section derives some comparative statics of our main results. To maximize transparency, our model is intentionally made simple with only two parameters besides information quality, both pertaining to the cost of investment: $k$, the firm's physical cost, and $c$, the manager's personal cost. The proposition below summarizes how they affect the screening of managers and the overall firm value:
\begin{proposition} \label{prop:comp stats}
Fixing any signal quality $\mathbf{q}>0$, $\theta_1(\mathbf{q})$, the minimal type in the managerial pool, is increasing in both the firm's investment cost $k$ and the manager's investment cost $c$. That is, $\partial \theta_1/\partial k >0$, and $\partial \theta_1/\partial c >0$. 

Suppose $\theta$ follows a uniform distribution. Firm value decreases in the manager's investment cost $c$ when the signal is either near perfectly informative (i.e., $q_1=q_0\rightarrow0$) or completely uninformative (i.e., $q_1=q_0=1/2$). That is, $\partial V(\mathbf{0})/\partial c < 0$ and $\partial V(\mathbf{1/2})/\partial c < 0$. Firm value  increases in the firm's investment cost $k$ when $k$ is low and decreases in $k$ when $k$ is high. That is, there exists some $k^*$ such that $\partial V(\mathbf{0})/\partial k > 0$ and $\partial V(\mathbf{1/2})/\partial k > 0$ when $k < k^*$, and $\partial V(\mathbf{0})/\partial k < 0$ and $\partial V(\mathbf{1/2})/\partial k < 0$ when $k > k^*$.
\end{proposition}

Proposition \ref{prop:comp stats} entails two sets of results. First, both the firm's investment cost $k$ and the manager's investment cost $c$ have the same \textit{positive} screening effect by increasing $\theta_1$. Intuitively, $\theta_1$ represents the lowest type the firm needs in order to invest upon receiving the good signal. Because both $k$ and $c$ represent the cost of investment but do not affect the gross return or the probability of success when signal quality $\mathbf{q}$ is given, the higher those costs, the higher the managerial type needed for investment to take place.\footnote{Based on the alternative intuition discussed in Section \ref{subsection:role of commitment}, a higher cost of investment mitigates the firm's lack of commitment problem by forcing the firm to withhold investment from a larger set of low-type managers even under the good signal.} 

However, while a higher $k$ or $c$ improves the screening, they both erode the firm's net return (profit) from investment. The second part of Proposition \ref{prop:comp stats} states that in the limiting cases when information becomes nearly perfectly informative or uninformative, the erosion of the firm's profit from investment dominates for all values of $c$. Interestingly, the effect of $k$ on the firm value is non-monotonic: when $k$ is low, the screening effect dominates, and firm value is increasing when $k$ becomes larger. When $k$ becomes sufficiently high, the erosion of profit due to a higher cost of investment dominates, and firm value is lower when $k$ becomes larger. In other words, given $\mathbf{q}$ and $c$, there exists an interior optimal level of $k$ that maximizes the firm value.


Proposition \ref{prop:comp stats} generates some interesting implications. If the firm has some control over the magnitude of the investment costs, the firm is better off driving down the manager's investment cost $c$ as low as possible. In Appendix \ref{appendix:MH}, we provide a micro-foundation for $c$ as the manager's cost of private effort necessary for the investment to succeed. Thus, the firm may be able to lower $c$ by improving internal operating efficiency, such as providing the manager with more skilled employees and reducing procedural requirements. Meanwhile, the firm does not always benefit from lowering the physical investment cost $k$. When such cost is low, a marginal increase of such cost may improve overall firm value by attracting on-average more capable managers through a positive screening effect.


\section{Conclusion}

Internal information is vital for the success of modern businesses. Conventional wisdom suggests that firms should design their internal information systems to be as precise as possible in order to make the most efficient operating decisions. In this paper, we highlight a novel mechanism through which a firm, when facing an adverse selection problem arising from unobservable managerial ability, may benefit from a noisy information system because it facilitates the screening of managers. The trade-off between more effective screening and more informed investment results in the \textit{ex ante} firm value to be non-monotonic in information quality, and a marginal improvement in information quality does not necessarily lead to an overall improvement in firm value. 

Our model can be extended in several directions. For simplicity, we consider a single representative firm facing a pool of managerial candidates. Introducing additional firms and allowing competition among them could produce new, insightful predictions. We also focus on adverse selection as the primary agency friction and consider a static model in which information is generated and used only once. Exploring the role of internal information in disciplining undesirable managerial actions could be fruitful, especially in a dynamic environment following the literature of dynamic moral hazard and monitoring (e.g., \citealp{PW2016}; \citealp*{VMS20}; \citealp{MS21}; etc.). In addition, to build a parsimonious model, we do not consider screening mechanisms other than the screening contract (i.e., the wage for a successful investment). In practice, firms could set up additional screening mechanisms to hire managers, which may complement and interact with the screening mechanism studied in our paper. We leave these interesting topics for future research.

\newpage \singlespacing
\appendix\setcounter{secnumdepth}{2} 

\begin{center}
{\LARGE \textbf{Appendix}}
\end{center}

\section{Extensions}
\subsection{Moral Hazard of the Manager} \label{appendix:MH}
In our baseline model, we assume that the firm is cash-constrained so that the firm can only make payments to the manager out of the cash flows generated by a successful investment. In this section we show that the compensation contract that the manager receives a bonus only upon project success can be micro-founded by the moral hazard problem of the manager. 

Specifically, consider an augmentation of our baseline model that, after the firm decides to invest, the manager has a binary effort choice: work, or shirk. If the manager chooses to work, the manager incurs a personal cost $c>0$ and the project may either succeed or fail depending on the underlying state as described in the baseline model. If the manager chooses to shirk, the manager incurs no cost but the project fails with certainty. The manager's effort choice is not observable to the firm, thus creating a moral hazard problem in addition to the adverse selection problem analyzed in our baseline model. 

Consider now the contract that the firm offers to the manager. The firm can offer a payment as soon as investment is made and make \textit{extra} payments if the project succeeds. Let $b$ denote the former payment and $w$ the latter (extra) one. In the unconditional investment zone (i.e., $\theta>\theta_0$), the incentive-compatible (IC) condition for the manager's effort choice is given by:
\begin{align}
	\theta w + b - c\geq b.
\end{align}
In the conditional investment zone (i.e., $ \theta\in(\theta_1, \theta_0]$), the IC condition is given by:
\begin{align}
	\theta (1-q_1)(w + b - c) + (1-\theta)q_0(b - c) \geq [\theta (1-q_1) + (1-\theta)q_0]b.
\end{align}
In both of the IC conditions, the left-hand side is the manager's expected utility upon working, while the right-hand side is the expected utility upon shirking. We assume the manager will exert the effort when indifferent. Rearranging the terms yields the following two IC conditions:
\begin{align}
	& \theta w - c \geq 0,\; \text{if } \theta>\theta_0, \label{Aeq:ICMH_u} \\
	& \theta (1-q_1)(w - c) - (1-\theta)q_0c \geq 0,\; \text{if } \theta\in(\theta_1, \theta_0].  \label{Aeq:ICMH_c}
\end{align}
The firm's contract $(b,w)$ must satisfy conditions (\ref{Aeq:ICMH_u}) and (\ref{Aeq:ICMH_c}) in addition to the incentive constraints for truthful reporting of managerial type in the baseline model, i.e., conditions (\ref{eq:IC_1}) and (\ref{eq:IC_0}). The following proposition summarizes the resulting equilibrium compensation contract and the implied managerial pool:

\begin{proposition} \label{prop:MH} 
Suppose the firm can make payment $b\geq 0$ for investment and extra payment $w\geq 0$ if the investment is successful, which requires the manager to exert effort at a cost $c>0$ that is not observable to the firm. The firm's incentive compatible contract that induces both truthful reporting and investment effort consists of $b=0$ and $w=w^* = c/(k+c)$. Managers are divided into three investment zones with boundaries $(\theta_0(\mathbf{q}), \theta_1(\mathbf{q}))$ that are identical to those characterized in Proposition \ref{prop:rent}. The firm's managerial pool is the same as that characterized in Proposition \ref{prop:screening} and the minimal type in the managerial pool $\theta_1(\mathbf{q})$ is increasing in the noisiness of the firm's signal. i.e., $\nabla \theta_1(\mathbf{q})>0$ for all $q_1, q_0$.
\end{proposition}
The proofs of these results are given later in Appendix \ref{appendix:proofs} but the intuition is as follows: when the firm can offer payment for a failed project, its investment conditions (\ref{eq:continuation condition1}) and (\ref{eq:continuation condition0}), as stated in the main text, can be re-written as:
\begin{align}
p_1(\theta^f_1)(1-w) = \hat k, \\
p_0(\theta^f_1)(1-w) = \hat k,
\end{align}
where $\hat k = k + b$. Similarly, conditions (\ref{eq:rent continuous1}) and (\ref{eq:rent continuous0}), which yield the thresholds for the conditional and unconditional investment zones in the main text, can be re-written as
\begin{align}  
\theta_1^m(1-q_1)(w - \hat c) - (1-\theta_1^m)q_0\hat c & = 0 ,\\
\theta_0^m(1-q_1)(w - \hat c) - (1-\theta_0^m)q_0\hat c & = \theta_0^mw - \hat c,
\end{align}
where $\hat c = c - b$. It is straightforward to see that these updated conditions in the model with the moral hazard problem are identical to the corresponding ones in the main text, i.e., conditions (\ref{eq:continuation condition1}),  (\ref{eq:continuation condition0}), (\ref{eq:rent continuous1}) and (\ref{eq:rent continuous0}), with a mere change of variables: $\hat k$ and $\hat c$ instead of $k$ and $c$. Thus, the  same analysis used in Section \ref{section:solution} applies and the extra wage for success compatible with truthful reporting is $w=\hat c/(\hat k + \hat c) = (c - b)/(\hat k + \hat c)$. Then, straightforward algebra shows that $w$ satisfy the IC constraints for the manager's effort (\ref{Aeq:ICMH_u}) and (\ref{Aeq:ICMH_c}) if and only if $b = 0$. Intuitively, making a payment upon project failure does not alleviate the agency frictions the firm is confronted with. However, to solicit the manager's effort, the firm must maintain a gap between the payment for project success and that for failure. Thus, in the equilibrium, the firm makes a payment $w^* = c/(k+c)$ if and only if the project succeeds and no payment if the project fails. The equilibrium compensation contract then divides the pool of managerial candidates into the same three zones as in the baseline model with the following boundaries:
\begin{align}
\theta_1(\mathbf{q}) & = \frac{q_0(\hat k+\hat c)}{(1-q_1)(1-\hat k-\hat c)+q_0(\hat k+\hat c)}  = \frac{q_0(k+c)}{(1-q_1)(1-k-c)+q_0(k+c)}  ,
\label{Aeq:theta1} \\
\theta_0(\mathbf{q}) & = \frac{(1-q_0)(\hat k+\hat c)}{q_1 + (1-q_1-q_0)(\hat k+\hat c)} = \frac{(1-q_0)(k+c)}{q_1 + (1-q_1-q_0)(k+c)}.
\label{Aeq:theta0}
\end{align}
Note that these thresholds are identical to the boundaries derived in the baseline model, i.e., conditions (\ref{eq:tilde theta1}) and (\ref{eq:tilde theta0}). Consequently, the screening effect of information quality (i.e., Proposition \ref{prop:screening}) and all subsequent analyses in our baseline model remain intact. 

\subsection{Positive Outside Option Value} \label{appendix:outside option}
This section explores a variation of the baseline model in which each manager has an outside option worth $R_{\min}>0$. We first present the case in which $R_{\min}$ is a constant. Then, we demonstrate the robustness of the model predictions when $R_{\min}$ is an increasing function of managerial type $\theta$. We also assume that when indifferent, managers prefer to be employed. This is without the loss of generality, as the results are virtually identical if we assume indifference implies unemployment instead.

Consider first the case in which $R_{\min}$ is a constant. For ease of exposition, we make the following technical assumption:
\begin{assumption} \label{assum:Rmin}
The firm's information noisiness $\{q_1, q_0\}$ is bounded by $\{\bar q_1, \bar q_0\}$ (i.e., $q_1\leq\bar q_1, q_0\leq\bar q_0$), where $\min[\bar q_1, \bar q_2]<1/2$. The manager's outside option $R_{\min}$ is a constant satisfying $R_{\min}<c\Delta$ where
\begin{align}
\Delta & \equiv \frac{(1-\bar q_1-\bar q_0)(1-k-c)}{(k+c)(1-\bar q_1-\bar q_0)+\bar q_1} >0.  \label{eq:Rmin low} 
\end{align}
\end{assumption}
In other words, we assume that the firm's signal is not completely uninformative,\footnote{More precisely, the only case ruled out is that both the good and the bad signals are completely uninformative (i.e., $q_1=q_2=1/2$).} and the manager's outside option value is not too high. As it will become clear later, this ensures that the minimal type of the managerial pool is within the conditional investment zone, where the screening effect of information quality manifests.

\begin{figure}[h]
\begin{center}
\includegraphics[width=0.9\textwidth]{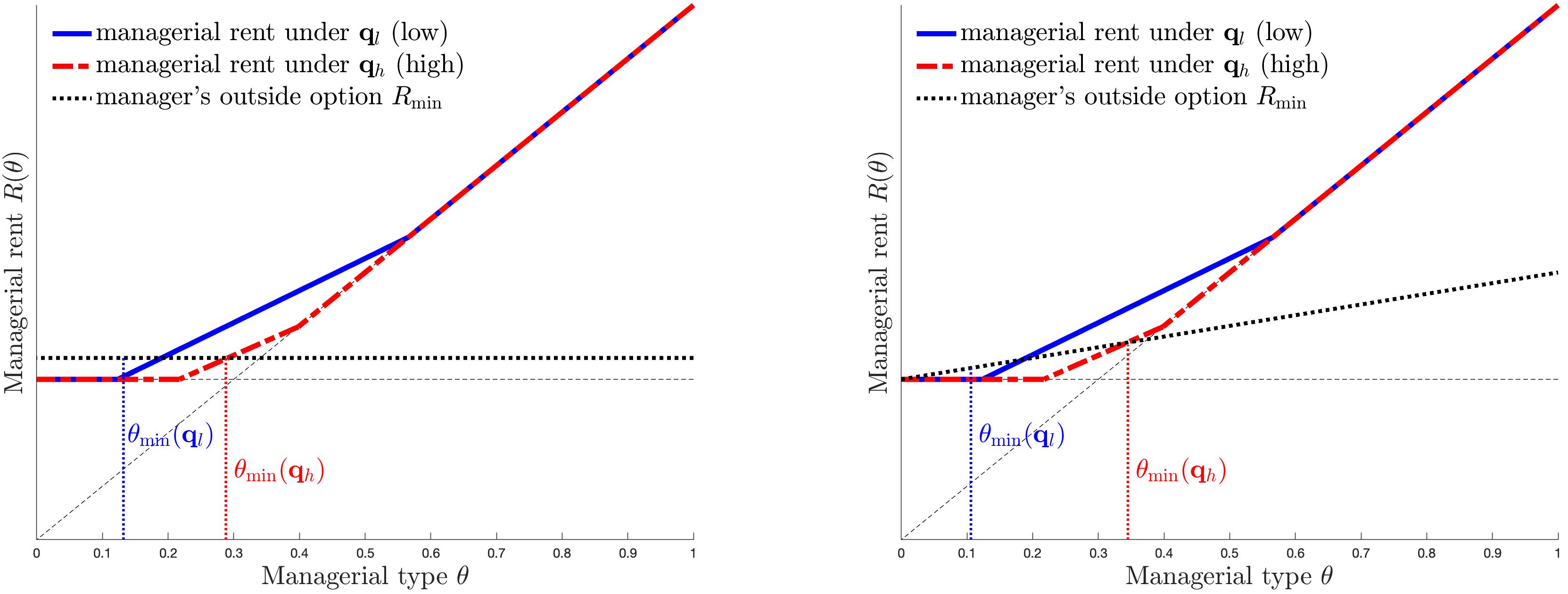}
\end{center}
\caption{{\footnotesize The left panel plots $\theta_{\min}(\mathbf{q})$, the minimal type in the firm's managerial pool, given by (\ref{Aeq:theta_min}), under different levels of information quality $\mathbf{q}_l$ and $\mathbf{q}_h$ (where $\mathbf{q}_l<\mathbf{q}_h$) when $R_{\min}$, the manager's outside option (black dotted line), is a positive constant. For better illustration, $q_1$ is set to equal $q_0$. The right panel plots plot $\theta_{\min}(\mathbf{q})$ under different levels of information quality $\mathbf{q}_l$ and $\mathbf{q}_h$ (where $\mathbf{q}_l<\mathbf{q}_h$) when $R_{\min}$ is a linearly increasing function of $\theta$.}}
\label{fig:R_min}
\end{figure}

Since the analysis in Section \ref{subsection:rent} does not depend on the value of outside options, all results in Proposition \ref{prop:rent} still apply. That is, the optimal wage is a constant given by (\ref{eq:wage}). The contract is incentive compatible: managers will report their types truthfully, and are divided into three investment zones with boundaries given by (\ref{eq:tilde theta1}) and (\ref{eq:tilde theta0}). However, because $R_{\min}>0$, the minimal type in the managerial pool, denoted by $\theta_{\min}$, no longer coincides with the left boundary of the conditional investment zone $\theta_1(\mathbf{q})$. Instead, $\theta_{\min}$ corresponds to the type of the manager whose rent from employment equals his outside option, or $R(\theta_{\min}) = R_{\min}$. The following proposition summarizes the properties of $\theta_{\min}$: 

\begin{proposition} \label{prop:R_min} 
Suppose each manager's outside option is worth a constant value $R_{\min}$. Given the firm's information quality $\mathbf{q}$, the contract derived in Proposition \ref{prop:rent} is optimal and incentive compatible. The firm's managerial pool is $\Theta = \{\theta:\theta\geq\theta_{\min}(q)\}$, where 
\begin{align} \label{Aeq:theta_min}
\theta_{\min}(\mathbf{q}) = \frac{(R_{\min} + q_0c)(k+c)}{\left((1-q_1)(1-k-c)+q_0(k+c)\right)c}.
\end{align}
If Assumption \ref{assum:Rmin} is satisfied, then $\theta_{\min}<\theta_0$, and $\nabla\theta_{\min}(\mathbf{q})>0$ for all $q_1, q_0$. That is, $\theta_{\min}$ is increasing in the noisiness of the firm's signal. 
\end{proposition}
The intuition behind this result is the same as that behind Proposition \ref{prop:screening} in the baseline model and can be illustrated using the left panel of Figure \ref{fig:R_min}. When $R_{\min}$ is a constant and not too high, the managerial pool consists of all the managers in the unconditional investment zone and some managers in the conditional investment zone. An increase in the noisiness of the signal narrows the conditional investment zone and lowers the marginal managerial rent in that zone. When information is more noisy, only managers with sufficiently high types can expect to receive rents commensurate with their outside options. Thus, a noisier signal achieves a better screening result by excluding more low-type candidates from the managerial pool.\footnote{If Assumption \ref{assum:Rmin} is not met, then $\nabla\theta_{\min}(\mathbf{q})>0$ when $q_1,q_0$ are small, and $\nabla\theta_{\min}(\mathbf{q})>0$ otherwise. In other words, the screening effect of noisier information still exists, just not for all levels of information quality.}

Our results are also qualitatively robust if $R_{\min}$ is an increasing function of $\theta$. That is, if the value of the manager's outside option increases with the manager's type. Because the analysis and the intuition of this case are very similar to that when $R_{\min}$ is constant, we omit the analysis in the interest of space and present the intuition graphically, using the right panel of Figure \ref{fig:R_min}. In this example, we assume that outside option value increases linearly in type, i.e., $R_{\min} = \rho\theta$ for some $\rho >0$. As the graph shows, $\theta_{\min}(\mathbf{q})$ is still increasing. That is, noisier information still achieves better screening.

Finally, because the distribution of managerial type is continuous, the results of both extensions illustrated in Figure \ref{fig:R_min} are virtually identical whether the managers prefer employment or not when they are indifferent (i.e., when their rent from employment equals the value of their outside options).

In short, the assumptions in the baseline model---that all managers have outside options worth zero and they favor unemployment when indifferent---are made for ease of exposition only. They do not materially affect the main result that noisier information can facilitate the screening for better managers.

\subsection{Alternative Matching Probability} \label{appendix:rand match}

In the baseline model, we assume that every manager in the managerial pool has an equal probability of being matched with the firm. This simplifies our analysis but is not crucial to our main results. Alternatively, we can relax this assumption and consider a model in which the probability of matching is captured by a distribution function $g_{\Theta}(\theta)$ with the properties that $g_{\Theta}(\theta)>0$ for all $\theta\in\Theta$, and $\int_{\Theta}g_{\Theta}(\theta)d\theta \equiv \int_{\Theta}dG_{\Theta}(\theta) = 1$. That is, regardless of the size of the managerial pool, every manager has some probability of being matched to the firm, and a match happens almost surely. Otherwise, we do not require any additional structure on the matching probability $g_{\Theta}(\theta)$. The matching used in the baseline model is essentially a special case in which $g_{\Theta}(\theta) = 1/(1-\theta_1)$, a constant regardless of the manager's type. However, it is possible to allow more general cases in which $g_{\Theta}(\theta)$ is increasing, decreasing, or non-monotonic in $\theta$; for instance, we may assume a higher-type manager is more likely to be matched with the firm. There exists a variety of rationales for the different shapes of the matching probabilities. 

Our main results remain under this modified assumption. First, because the manager's choice of whether to join the managerial pool depends only on their expected utility from being hired, the boundaries of the conditional investment zone $\theta_1(\mathbf{q})$ and $\theta_0(\mathbf{q})$ as well as the observations in Propositions \ref{prop:rent} and \ref{prop:screening} are unaffected. Expected firm value $V(\mathbf{q})$, as defined in (\ref{eq:V(q)_equilibrium}), now becomes 
\begin{align}
V(\mathbf{q}) & = \left(\frac{1}{1-F(\theta_1)} \right)\left[\int^{\theta_0 }_{\theta_1}v_c(\theta)g_{\Theta}(\theta)dF(\theta) +\int^{1}_{\theta_0}v_u(\theta)g_{\Theta}(\theta)dF(\theta) \right] \\
& = \left(\frac{1}{1-F(\theta_1)} \right)\left[\int^{\theta_0 }_{\theta_1}v_c(\theta)d\tilde F(\theta) +\int^{1}_{\theta_0}v_u(\theta)d\tilde F(\theta) \right], \label{Aeq:V(q)_alt match} 
\end{align}
where $\tilde F(\theta) = F(\theta)G_{\Theta}(\theta)$. Thus, the implications of the model under the modified matching probability can be replicated by those from a model with random matching but a different distribution of the managers. In other words, because our model does not place any restriction on the distribution of the managers' types, $F(\theta)$, the assumption that all types of managers inside the managerial pool have an equal probability of being matched with a firm does not come with a loss of generality.

\section{Proofs} \label{appendix:proofs}

\subsection{Proof of Proposition \ref{prop:rent}}

The proof is divided into two parts. First, we solve the equilibrium wage and the boundaries of the conditional investment zones. Then, we show that a constant wage is indeed optimal.

\bigskip

\noindent \textbf{Part I}: Define 
\begin{align}
\mu_0(\theta) & = \frac{(1-\theta)(1-q_0)}{\theta q_1},
\end{align}
as the likelihood ratio between failure and success given a bad signal, and 
\begin{align}
\mu_1(\theta) & = \frac{(1-\theta)q_0}{\theta (1-q_1)},
\end{align}
the likelihood ratio between failure and success given a good signal. Then, 
\begin{align}
p_0(\theta) & = \frac{\theta q_1}{\theta q_1 + (1 - \theta)(1-q_0)} = \frac{1}{%
1+\mu_0(\theta)} ,\\
p_1(\theta) & = \frac{\theta (1-q_1)}{\theta (1-q_1) + (1 - \theta)q_0} = \frac{1}{%
1+\mu_1(\theta)}.
\end{align}
Using these notations, (\ref{eq:continuation condition1}) and (\ref%
{eq:continuation condition0}) imply 
\begin{align}
w & = 1 - (1+\mu_1(\theta^f_1))k = 1 - (1+\mu_0(\theta^f_0))k
\label{Aeq:w=1+mu}.
\end{align}
Combining them with (\ref{eq:rent continuous1}) and (\ref{eq:rent
continuous0}), and imposing the incentive compatibility condition $%
\theta^f_s=\theta^m_s$ ($s=0,1$) yield 
\begin{align}
w & = w^* \equiv \frac{c}{k + c},
\end{align}
which, substituted into (\ref{Aeq:w=1+mu}), implies 
\begin{align}
1 + \mu_1(\theta_1) = 1 + \mu_0(\theta_0) = \frac{1}{k + c},
\end{align}
which yields (\ref{eq:tilde theta1}) and (\ref{eq:tilde theta0}). It is easy to verify that $\theta_1>0$, $\theta_0(q)<1$, and $\theta_1<\theta_0$ always when $0<q_1, q_2<1/2$. i.e., the three investment zones exist for all $q_1,q_2$. When $q_1=q_2 = 1/2$, $\theta_1(1/2)=\theta_0(1/2)=k+c<1$, i.e., the conditional investment zone disappears. $\square$

\bigskip

\noindent \textbf{Part II}: $w^*$ must be a constant under the optimal screening contract. First, wage must be independent of type within each investment zone. If not, then a manager can report his type to be the one that earns him the highest wage while not altering his investment zone. Thus there can be at most two wages: $w(\hat\theta) = w_c$ if $\hat\theta<\theta_0$, and $w(\hat\theta) = w_u$ if $\hat\theta\geq\theta_0$, where $\theta_0$ is the boundary between the conditional investment zone and the unconditional investment zone satisfying the firm's ex-post optimal investment condition: $p_0(\theta_0)(1-w_u)=k$. Suppose $w_c<w_u$. Then, because $p_0(\theta)$ decreases smoothly in $\theta$, there must exist $\dot\theta<\theta_0$ such that $p_0(\theta)(1-w_c)>k$ for all $\theta\in(\dot\theta,\theta_0)$, contradictory to the fact that $\theta_0$ is the boundary of the conditional investment zone. Therefore it must be that $w_c\geq w_u$. However, given $w_u$, let $R^i(\theta)$, $\theta^i_0$, and $\theta^i_1$ ($i=1,2$) be the managerial rent and the boundaries of the conditional investment zones for $w^i_c$, with $w^1_c<w^2_c$. Then $\theta^1_1>\theta^2_1$ (i.e., a better screening result), and $R^1(\theta)< R^2(\theta)$ for all $\theta\in(\theta^1_1,\theta^2_0)$ (i.e., lower rents for managers in the conditional investment zone). Thus $w^1_c$ is strictly preferred to $w^2_c$ by the firm. Therefore, it must be that $w_c=w_u=w$ for all managers under the optimal screening contract. $\square$

\subsection{Proof of Propositions \ref{prop:screening} and \ref{prop:conservatism}}

The proof is straightforward. First, from (\ref{eq:tilde theta1})
\begin{align}
\theta_1 & = \frac{1}{1 + \left(\frac{1-q_1}{q_0}\right)\gamma}, 
\end{align}
where $\gamma = \frac{1}{k+c} - 1>0$. Clearly, $\partial\theta_1/\partial q_i >0$ for $i\in\{1,0\}$. Next, differentiating $\theta_1(q,\lambda)$ from (\ref{eq:theta_1_conservatism}) with respect to $q$ and $\lambda$ yields: 
\begin{align}
\frac{\partial \theta_1(q,\lambda)}{\partial q} & = \frac{k+c}{(1-q-\lambda)(1-k-c)+(q-\lambda)} - \frac{(q-\lambda)(k+c)^2}{[(1-q-\lambda)(1-k-c)+(q-\lambda)]^2}  \notag \\
& = \frac{(k+c)[(1-2q)(1-k-c) + (q-\lambda)]}{[(1-q-\lambda)(1-k-c)+(q-\lambda)]^2} > 0. \\
\frac{\partial \theta_1(q,\lambda)}{\partial \lambda} & = -\frac{k+c}{(1-q-\lambda)(1-k-c)+(q-\lambda)} + \frac{(q-\lambda)(k+c)(2-k-c)}{[(1-q-\lambda)(1-k-c)+(q-\lambda)]^2}  \notag \\
& = -\frac{(k+c)(1-2q)(1-k-c)}{[(1-q-\lambda)(1-k-c)+(q-\lambda)]^2} < 0.
\end{align}
because $1-2q>0$ and $q>\lambda$. $\square$

\subsection{Proof of Propositions \ref{prop:Vprime} and \ref{prop:comp stats}}
The proof is organized into three parts.

\bigskip
\noindent \textbf{Part I}: Substituting (\ref{eq:tilde theta1}) into $v_c(\theta)$ implies that $v_c(\theta_1)=0$. Then, $v_c(\theta)>0$ for all $\theta\in(\theta_1,\theta_0)$ because
\[
v'_c(\theta) = (1-q_1)(1-w^*-k) + q_0k >0 .
\]
Substituting (\ref{eq:tilde theta0}) into $v_u(\theta)$ implies that $v_u(\theta_0)=v_c(\theta_0)>0$. Then, $v'_u(\theta) = 1-w^*>0$ implies that $v_u(\theta)>0$ for all  $\theta\in(\theta_0,1)$. Therefore, $V(\mathbf{q})>0$ for all $\mathbf{q}$.

\bigskip
\noindent \textbf{Part II}: First, from (\ref{eq:tilde theta1}), $q_0 \rightarrow 0$ implies $\theta_1 \rightarrow 0$ for all $q_1$. Thus,
\begin{align}
\left.\frac{\partial V(q_1,q_0)}{\partial q_1}\right|_{q_0\rightarrow 0} = \left(\frac{1}{1-F(\theta_1)} \right)\left[\int^{\theta_0 }_{0}\left(\frac{\partial v_c}{\partial q_1}\right)dF(\theta)\right] <0. 
\end{align}
Next, from (\ref{eq:tilde theta0}), $q_1 \rightarrow 0$ implies $\theta_0 \rightarrow 1$ and 
\[
\theta_1 \rightarrow 1 - \frac{\gamma}{q_0 + \gamma}
\]
for all $q_0$. Thus,
\begin{align}
\left.\frac{\partial V(q_1,q_0)}{\partial q_0}\right|_{q_1\rightarrow 0}= \left(\frac{1}{1-F(\theta_1)}\right)\int_{\theta_1}^1 \left(\frac{g(\theta)}{\phi}\right)d\theta,
\end{align}
where
\[
g(\theta)\equiv \theta(1-w^*-k + q_0k + \phi k) - (\phi+q_0)k, 
\]
and
\[
\phi \equiv \frac{(q_0 + \gamma)^2}{\gamma} >0.
\]
Define $\theta_m$ such that $g(\theta_m)=0$. Because
\[
\frac{(\phi+q_0)k}{1-w^*-k+(1+q_0)k} = \frac{\phi + q_0}{\gamma + q_0 + \phi} > \frac{\gamma }{2(q_0 + \gamma)}, 
\]
we have $\theta_m > (1-\theta_1)/2$. Thus, $\int_{\theta_1}^1g(\theta)d\theta<0$, implying that $\partial V(0,q_0)/\partial q_0<0$. Finally, $\theta_0 \rightarrow \theta_1$ when $q_1 = q_0 = 1/2$. Thus,
\[
\nabla V\left(\mathbf{\frac{1}{2}}\right) = V\left(\mathbf{\frac{1}{2}}\right)\left(\frac{f(\theta_1)}{1-F(\theta_1)}\right)\nabla V\theta_1\left(\mathbf{\frac{1}{2}}\right) >0.
\]
Because $\nabla V(\mathbf{q})$ is continuous, there exist $0<\underline q<\overline q<1/2$ such that $\nabla V(\mathbf{q}) < 0$ if $q_1 <\underline q$ or $q_0<\underline q$, and $\nabla V(\mathbf{q}) > 0$ if $q_1 >\underline q$ and $q_0 > \underline q$.

\bigskip
\noindent \textbf{Part III}: The results that $\partial \theta_1/\partial k > 0$ and $\partial \theta_1/\partial c > 0$ are obvious. 
If $\theta$ follows a uniform distribution, then
\begin{align*}
V(\mathbf{0}) & = \int_0^1\theta(1-w^*-k)d\theta = \frac{k}{2}\left(\frac{1-k-c}{k+c}\right) = \frac{1}{1-k-c}\int^1_{k+c}(\theta w^* - k)d\theta  = V\left(\mathbf{\frac{1}{2}}\right) .
\end{align*}
Clearly, $\partial V(\mathbf{0})/\partial c < 0$ and $\partial V(\mathbf{1/2})/\partial c < 0$. Define
\[
h(k) = \ln\left[k\left(\frac{1-k-c}{k+c}\right)\right] = \ln k + \ln(1-k-c) - \ln(k+c).
\]
Then
\begin{align}
h'(k) & = \frac{1}{k} - \frac{1}{1-k-c} - \frac{1}{k+c},
\end{align}
with $\lim_{k\rightarrow 0}h'(k) > 0$ and $\lim_{k\rightarrow 1-c}h'(k) < 0$. Since
\begin{align}
h''(k) & = -\frac{1}{k^2} -\frac{1}{(1-k-c)^2} + \frac{1}{(k+c)^2} <0,
\end{align}
there exists $k^*$ such that $h'(k)>0$ for all $0<k<k^*$ and $h'(k)<0$ for all $k^*<k<1-c$. Consequently, $\partial V(\mathbf{0})/\partial k >0$ and  $\partial V(\mathbf{1/2})/\partial k >0$ for all $0<k<k^*$, while $\partial V(\mathbf{0})/\partial k < 0$ and  $\partial V(\mathbf{1/2})/\partial k < 0$ for all $k^*<k<1-c$.  $\square$

\subsection{Proof of Propositions \ref{prop:MH}}
Following the change of variable $\hat k = k + b$ and $\hat c = c - b$, the IC constraints for truthful reporting are given by (\ref{eq:IC_1}) and (\ref{eq:IC_0}) yield $w = \hat c/(k+c)=(c-b)/(k+c)$. The boundaries of the conditional investment zone are given by
\begin{align}
\theta_1(\mathbf{q}) & = \frac{q_0(\hat k+\hat c)}{(1-q_1)(1-\hat k-\hat c)+q_0(\hat k+\hat c)}  = \frac{q_0(k+c)}{(1-q_1)(1-k-c)+q_0(k+c)}, \\
\theta_0(\mathbf{q}) & = \frac{(1-q_0)(\hat k+\hat c)}{q_1 + (1-q_1-q_0)(\hat k+\hat c)} = \frac{(1-q_0)(k+c)}{q_1 + (1-q_1-q_0)(k+c)}.
\end{align}
The incentive constraints for investment effort (\ref{Aeq:ICMH_u}) and (\ref{Aeq:ICMH_c}) becomes
\begin{align} \label{Aeq:MH_w_condition}
w & \geq \max\left\{\frac{c}{\theta_0}, \; \left(1+\frac{(1-\theta_1)q_0}{\theta_1(1-q_1)}\right)c\right\}.
\end{align}
Because $(q_1, q_0) < 1/2$,
\begin{align}
\frac{1}{\theta_0} - 1 = \frac{(1-k-c)q_1}{(k+c)(1-q_0)}< \frac{1-k-c}{k+c}= \frac{(1-\theta_1)q_0}{\theta_1(1-q_1)}.
\end{align}	
Thus, (\ref{Aeq:MH_w_condition}) requires
\begin{align}
w = \frac{c-b}{k+c} \geq \left(1+\frac{1-k-c}{k+c}\right)c = \frac{c}{k+c},
\end{align}
which is satisfied if and only if $b = 0$. Thus, $w = w^* = c/(k+c)$. $\theta_1(\mathbf{q})$ and $\theta_0(\mathbf{q})$ are identical to those given in (\ref{eq:tilde theta1}) and (\ref{eq:tilde theta0}), respectively, and the corresponding properties characterized in Propositions \ref{prop:rent} and \ref{prop:screening} follow through. $\square$

\subsection{Proof of Propositions \ref{prop:R_min}}
The proof of Proposition \ref{prop:rent} does not involve the value of $R_{\min}$ and thus still applies. That is, the contract characterized in Proposition \ref{prop:rent} is still optimal and incentive compatible. Given that $R(\theta)$ is increasing in $\theta$, $\Theta = \{\theta:\theta\geq\theta_{\min}(\mathbf{q})\}$, where $R(\theta_{\min}(\mathbf{q})) = R_{\min}$ yields (\ref{Aeq:theta_min}). If Assumption \ref{assum:Rmin} holds, then 
\begin{align}
\theta_{\min}(\mathbf{q}) < \frac{(\Delta+ q_0)(k+c)}{(1-q_1)(1-k-c)+q_0(k+c)} < \theta_0 ,
\end{align}
for all $q<\bar q$. Differentiating $\theta_{\min}(\mathbf{q})$ yields
\begin{align}
\nabla \theta_{\min}(\mathbf{q}) = 
\begin{bmatrix}
\frac{\theta_{\min}(\mathbf{q})(1-k-c)}{(1-q_1)(1-k-c)+q_0(k+c)} \\
\frac{(1-\theta_{\min}(\mathbf{q}))(k+c)}{(1-q_1)(1-k-c)+q_0(k+c)}
\end{bmatrix}
 > 0 .
\end{align}
$\square$

\clearpage
\newpage \singlespacing
\bibliographystyle{chicago}
\bibliography{FWWZ}

\end{document}